\documentclass[aps,rmp,twocolumn,floatfix]{revtex4}
\pdfoutput=1
\usepackage{graphicx}
\usepackage{epstopdf}
\usepackage{amssymb}

\newcommand{\ha}{H\ensuremath{\alpha}}
\newcommand{\hb}{H\ensuremath{\beta}}
\newcommand{\hii}{H~{\sc ii}}
\newcommand{\hi}{H~{\sc i}}
\newcommand{\kms}{km~s\ensuremath{^{-1}}}
\newcommand{\vlsr}{\ensuremath{v_{\mathrm{LSR}}}}
\newcommand{\iha}{\ensuremath{I_{\mathrm{H}\alpha}}}
\newcommand{\nhi}{\ensuremath{N_{\mathrm{H\;I}}}}
\newcommand{\sii}{[S~{\sc ii}]}
\newcommand{\nii}{[N~{\sc ii}]}
\newcommand{\hei}{He~{\sc i}}
\newcommand{\oi}{[O~{\sc i}]}
\newcommand{\oii}{[O~{\sc ii}]}
\newcommand{\oiii}{[O~{\sc iii}]}
\newcommand{\degree}{\ensuremath{^{\circ}}}

\begin{document}

\title{The warm ionized medium in spiral galaxies}

\author{L.M. Haffner}
\affiliation{Department of Astronomy, University of Wisconsin--Madison, USA}

\author{R.-J. Dettmar}
\affiliation{Astronomisches Institut, Ruhr-University Bochum, Germany}

\author{J.E. Beckman}
\affiliation{Instituto de Astrof\'\i sica de Canarias, Tenerife, Spain}
\affiliation{Consejo Superior de Investigaciones Cient\'\i ficas, Spain}

\author{K. Wood}
\affiliation{University of St Andrews, Scotland}

\author{J.D. Slavin}
\affiliation{Harvard-Smithsonian Center for Astrophysics,  Cambridge, MA, USA}

\author{C. Giammanco}
\affiliation{Instituto de Astrof\'\i sica de Canarias, Tenerife, Spain}

\author{G.J. Madsen}
\affiliation{School of Physics, The University of Sydney, NSW 2006, 
Australia}

\author{A. Zurita}
\affiliation{Dept. de F\'\i sica Te\'orica y del Cosmos, U. de Granada, Granada, Spain}

\author{R.J. Reynolds}
\affiliation{Department of Astronomy, University of Wisconsin--Madison,\ USA}

\begin{abstract}
This article reviews observations and models of the diffuse ionized gas 
that permeates the disk and halo of our Galaxy and others. It was inspired 
by a series of invited talks presented during an afternoon scientific 
session of the 65th birthday celebration for Professor Carl Heiles held at 
Arecibo Observatory in August 2004.  This review is in recognition of 
Carl's long standing interest in and advocacy for studies of the 
\emph{ionized} as well as the neutral components of the interstellar 
medium.
\end{abstract}

\maketitle
\tableofcontents

\section{Introduction}
\label{intro}

Composed primarily of hydrogen (91\% by number) and helium (9\%), with 
trace amounts (0.1\%) of heavier elements, the interstellar medium plays a 
vital role in the cycle of stellar birth and death and galactic evolution.  
Not only do the properties of the interstellar medium govern the 
formation of new stars, but through their radiation and the matter and 
kinetic energy from their outflows and supernovae, these stars in turn 
determine the properties of the interstellar medium from which the next 
generation of stars will be born. One of these feedback processes, the 
subject of this review, is the large-scale ionization of the medium by the 
youngest and most luminous stars, the O~stars.  Even though they are 
located near the galactic midplane in rare, isolated regions of star 
formation and often surrounded by opaque clouds of neutral hydrogen, the 
Lyman continuum radiation from these hot stars is somehow able to 
propagate large distances through the disk and into the galaxy's halo to 
produce extensive ionization of the interstellar hydrogen. The study
of this wide-spread plasma has impacted our understanding of the dynamic 
interstellar processes occurring in galaxies.  

This area of study began more than four decades ago, when \citet{HE63} 
proposed to a skeptical astronomical community the existence of an 
extensive layer of warm (10$^4$~K), low density (10$^{-1}$~cm$^{-3}$) 
ionized hydrogen surrounding the plane of our Galaxy and having a power 
requirement comparable to the ionizing luminosity of the Galaxy's O and B 
stars.  Their conclusion was based upon their discovery of a free-free 
absorption signature in the observations of the Galactic synchrotron 
background at frequencies between 1.5 and 10 MHz carried out by radio 
astronomy pioneers Grote Reber and G. R. A. Ellis at Hobart, Tasmania 
\citep{RE56, EWB62}.  The idea that a significant fraction of the Galaxy's 
ionizing photons, produced primarily by rare, massive stars residing near 
the Galactic midplane, traveled hundreds of parsecs throughout the disk to 
produce wide-spread ionization conflicted with the traditional picture in 
which the neutral atomic hydrogen (the primary component of the medium by 
mass and opaque to hydrogen ionizing radiation) filled much of the 
interstellar volume.  However, less than a decade later, the dispersion of 
radio signals from newly discovered pulsars \citep{HBP68, Guelin74} plus 
the detection of faint optical emission lines from the diffuse 
interstellar medium \citep{Reynolds71, RSR73} firmly established warm 
ionized hydrogen as a major, wide-spread component of our Galaxy's 
interstellar medium. Two decades later, deep H$\alpha$ imaging with CCD 
detectors revealed that similar warm plasmas also permeate the disks and 
halos of other galaxies \citep{RKH90, Dettmar90}.

The large mass, thickness, and power associated with this ionized layer 
has modified our understanding of the composition and structure of the 
interstellar medium and the distribution of Lyman continuum radiation 
within galaxies.  Its weight is a major source of interstellar pressure at 
the midplane \citep{BC90}, and it could be the dominant state of the 
interstellar medium 1000 pc above the midplane \citep{Reynolds91a}.  An 
accurate understanding of interstellar matter and processes thus requires 
as thorough a knowledge of this diffuse ionized gas as the other principal 
components of the medium.  For a general survey of the interstellar 
medium, we refer the reader to \emph{The Interstellar Environment of Our 
Galaxy} by Katia \citet{Ferriere01} and \emph{The Three Phase Interstellar 
Medium Revisited} by Donald \citet{Cox05}.  Brief review articles on a 
variety of current interstellar medium topics can be found in \emph{How 
Does the Galaxy Work?}, a conference proceedings edited by \citet{APF04}.

Although the nature of this low density plasma is not yet fully 
understood, significant progress has been made in characterizing its 
properties and the source of its ionization.  In \S \ref{sec:diving}, we 
present a brief account of observations of the warm ionized medium (WIM) 
in our Galaxy, the Milky Way, including results on the physical conditions 
in the WIM. In \S \ref{sec:edgeon} and \S \ref{sec:faceon}, we review 
progress on understanding this gas and its close connection to star 
formation activity and O~stars through studies of other galaxies. O star 
photoionization models in a clumpy interstellar medium are discussed in \S 
\ref{sec:clumpy}, with a discussion of supplemental radiation from 
hot-cool gas interfaces in \S \ref{sec:interface}. A list of some 
unanswered questions and future challenges is presented in \S 
\ref{sec:questions}. Throughout, we keep the convention that WIM refers to 
the warm ionized medium in our Galaxy and DIG refers to this diffuse 
ionized gas in other galaxies.  Also, we use $\mathrm{H}^{+}$ to refer to 
the ionized hydrogen in the low density, wide-spread WIM/DIG and \hii\ to 
denote the gas in the localized, higher density, ``classical \hii\ 
regions'' immediately surrounding hot stars. For reference, a list of 
acronyms and terms with explanations is provided at the end of this paper.

\section{View from the inside: diving into the physics of the warm ionized medium in our galaxy}
\label{sec:diving}

\subsection{Basic characteristics of the WIM and diagnostic tools}
\label{sec:basic}

Our location within the disk of the Galaxy provides an opportunity to 
explore close up and in detail the distribution and physical properties of 
this ionized medium, including its ionization state and temperature. The 
basic features of the WIM are not very different from those first proposed 
by \citet{HE63}. Temperatures range from about 6000~K to 10~000~K, and in 
the solar neighborhood, its average hydrogen ionization rate is 
approximately $4 \times 10^6$ s$^{-1}$ within a one cm$^{2}$ column 
extending perpendicular through the Galactic disk, about 1/8th that 
available from stellar ionizing photons \citep[e.g.,][]{MRH06, 
Reynolds84}.  The amount of ionization increases toward the Galactic 
center \citep[e.g.,][]{MR05}. In other galaxies, the average ionization 
rate is observed to be about 1/2 that available from the stars (see \S 
\ref{sec:faceon}). 

Two fundamental parameters of ionized gas along any line of sight, $s$, are 
the dispersion measures (DM $\equiv \int n_e\,ds$), derived from pulsar observations, and the
emission measure (EM $\equiv \int n_e n_{\mathrm{H}^+}\,ds \approx \int n_{e}^2\,ds$), derived 
from the intensity of the hydrogen Balmer-alpha (\ha) recombination line or from the amount of free-free (Bremsstrahlung) emission or absorption. Comparison of these measurements along common lines of sight indicate that the $\mathrm{H}^{+}$ is clumped into 
regions having an average electron density, n$_{e} = $ 0.03--0.08 cm$^{-3}$,
and filling a fraction, $f \approx$ 0.4--0.2, of the volume within a 2000--3000 pc thick layer about the Galactic midplane \citep{Hill08, 
Reynolds91b}.  Data also suggest that the filling fraction increases from 
$f \sim 0.1$ at the midplane to $f >$ 0.3--0.4 at $|z|$ = 1000~pc 
\citep{KH87, Reynolds91b, GMC08, BMM08}. The large, 1000--1800~pc scale 
height, significantly larger than that of the neutral hydrogen layer, has 
been deduced from both pulsar observations \citep{Reynolds89, GMC08} and 
from the rate of decrease in the \ha\ intensity with increasing Galactic 
latitude for the gas associated with the Perseus spiral arm 
\citep[e.g.,][]{HRT99}. The WIM accounts for 90\% or more of the 
ionized hydrogen within the interstellar medium, and along lines of sight 
at high Galactic latitude (i.e., away from the Galactic midplane), the 
column density of the $\mathrm{H}^{+}$ is approximately 1/3 that of the 
neutral hydrogen \citep{Reynolds91a}.

\begin{figure}[tbp]
\begin{center}
\resizebox{\linewidth}{!}{\includegraphics{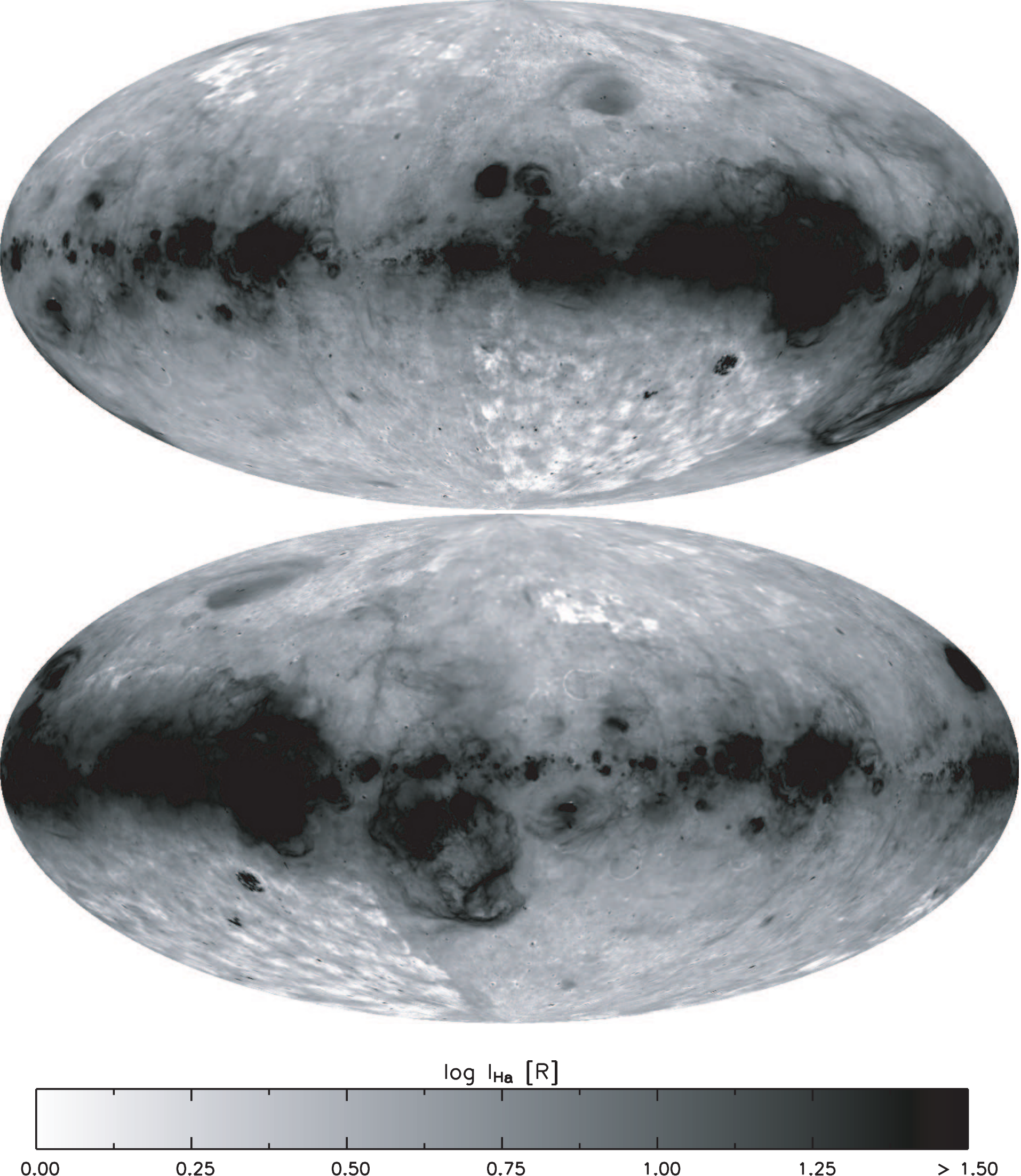}}
\caption{Composite all-sky velocity-integrated \ha\ map. Data from VTSS, 
SHASSA, and the WHAM northern sky survey (WHAM-NSS) have been combined to 
produce these very deep (EM $\agt 1$ cm$^{-6}$ pc) emission maps. The two Aitoff-Hammer projections 
are centered at  (\emph{top}) $\ell=0\degree$and (\emph{bottom}) $\ell=180\degree$. Areas covered by the imaging surveys have arc-minute
resolution while those only surveyed by WHAM have one-degree resolution.
Emission is predominantly from the Galaxy, although the imaging surveys
can contain other bright, extended sources from the local universe
($|\vlsr| \alt 500$ \kms; LMC, SMC, M31, etc.).  Adopted from \citealp{Finkbeiner03}.}
\label{fig:allsky}
\end{center}
\end{figure}

Although originally detected by radio observations, subsequent 
developments in high-throughput Fabry-Perot spectroscopy and CCD imaging 
techniques have demonstrated that the primary source of information about 
the distribution, kinematics, and other physical properties of the WIM is 
through the detection and study of faint interstellar emission lines at 
optical wavelengths. For example, the distribution of the $\mathrm{H}^{+}$ 
is revealed by its interstellar \ha\ ($\lambda6563$) recombination line 
emission, which covers the sky. Several deep \ha\ surveys have given us 
our first detailed view of the distribution and kinematics of this gas. 
Through CCD imaging, \citet{VTSS} has provided partial coverage of the 
total \ha\ intensity in the northern sky, while \citet{SHASSA} covered the 
southern sky at arcminute resolution and sensitivities around 1 R 
($10^{6}/4\pi$~photons cm$^{-2}$ s$^{-1}$ sr$^{-1}$; EM = 2.25~cm$^{-6}$ 
pc at 8000~K). The Wisconsin H-Alpha Mapper (WHAM) carried out a velocity 
resolved survey of the northern sky (dec. $> 30\degree$) at 12 \kms\ spectral 
resolution, $1\degree$ spatial resolution, and $\sim 0.1$ R sensitivity 
\citep{WHAMNSS, TuftePhD}. The WHAM maps show WIM emission in virtually 
every beam, with faint loops, filaments, and blobs of emission superposed 
on a more diffuse background. \citet{Finkbeiner03} has combined these 
three surveys to form a composite all-sky view of the velocity-integrated 
\ha\ (Fig.~\ref{fig:allsky}).

With these new maps and methods for detecting faint emission lines, we are 
now able to investigate the physical conditions of the WIM, its 
relationship to other components of the interstellar medium, and to 
sources of ionization and heating within the Galaxy. In particular, 
standard nebular line diagnostic techniques can now be employed to examine 
the physical conditions in the gas. In the low density ($\sim 10^{-1}$ 
cm$^{-3}$) environment of the WIM, the collisional excitation of an ion to 
a metastable state 2--3~eV above ground by the thermal ($\sim 
10^{4}$~K) electrons is followed by the decay back to the ground state via 
a ``forbidden'' optical transition.  Specifically, the ion's excitation 
rate r$_{i} \propto$ $n_i$ $n_e$ $T_{e}^{-0.5} \exp(-E/kT_e$), 
where 
$n_i$ and $n_e$ are the volume densities of the ions and electrons, 
respectively, $T_e$ is the electron temperature, and $E$ is the energy of 
the metastable state above ground.  Because thermal equilibrium between 
electrons and ions is very rapid, the temperature of the ions $T_i$ = 
$T_e$ \citep[e.g.,][]{Spitzer78}.  Thus a variation in the photon 
emissivity of a forbidden line from one direction to the next traces 
variations in the temperature, 
density, and abundance of the ion.  The effects of density variations can 
be eliminated by dividing the forbidden line intensity by the 
H-recombination line intensity, both of which are proportional to the 
product $n_i\ n_e$.  From the intensities of lines from a number of 
different ions and atoms, it has been possible to study separately 
variations in the temperature and the ionization state within the emitting 
gas.  For many years, these diagnostic techniques have been applied to a 
variety of astrophysical plasmas \citep[see, e.g.,][for in depth 
discussions] {o89, OF06, DS03, fer03, DN79}, but only more recently has it 
been possible to use them to study the much fainter WIM emission lines.

For example, in the WIM, the forbidden lines \sii\ $\lambda6716$ and \nii\ 
$\lambda6584$ are found to have intensities with respect to \ha\ that 
range from a few tenths to unity or higher, significantly larger than what 
is observed for the bright, classical emission nebulae (i.e., \hii\ 
regions) immediately surrounding O stars. This implies that the physical 
conditions in the WIM differ significantly from conditions in classical 
\hii\ regions. In addition, because their intensities are comparable to 
H$\alpha$, it has been possible to map these lines over large parts of the 
sky \citep[e.g.,][]{MRH06}.  Other lines, such as \nii\ $\lambda5755$, 
\hei\ $\lambda5876$, \oiii\ $\lambda 5007$, and \oi\ $\lambda6300$, are 
much fainter and have been studied only in a few select directions. These 
observations have helped to characterize the ionization and temperature of 
the WIM as well as other extended ionized regions of the Galaxy.  Results 
reveal that not only are the temperature and ionization conditions of the 
WIM significantly different from the conditions in classical O star \hii\ 
regions, but that the conditions within the WIM itself vary considerably 
from one direction to the next and even along a single line of sight 
\citep{MRH06}.

\subsection{Ionization state}
\label{sec:ionization}

The strength of the ionizing radiation field responsible for the WIM can 
be probed by measuring the \ha\ surface brightness of neutral hydrogen 
(\hi) clouds and by measuring the hydrogen ionization fraction 
$\mathrm{H}^{+}/\mathrm{H}$ within the WIM.  \citet{Field75} pointed out 
that an \hi\ cloud immersed in an ionizing radiation field will have a 
skin of $\mathrm{H}^{+}$ with an emission measure that is directly 
proportional to the incident photon flux. Using this fact, \citet{RTK95} 
found that an interstellar Lyman continuum flux $4\pi$ J $\approx 2 \times 
10^6$ photons cm$^{-2}$ s$^{-1}$ could account for most of the WIM's 
ionization. This flux implies an ionizing photon density to electron 
density ratio (the ionization parameter) of $10^{-4}$ to $10^{-3}$, which 
is one to two orders of magnitude smaller than the ionization parameter in 
classical O star \hii\ regions. However, values in this range still imply 
that the hydrogen is nearly fully ionized within the WIM.  This is confirmed 
by more direct measurements of the hydrogen ionization fraction from the 
detection of neutral oxygen emission.

In theory, directly measuring the degree of H-ionization within warm, 
ionized gas is simply a matter of observing the \oi\ $\lambda6300$ 
emission line, which is produced by collisions of neutral oxygen with 
thermal electrons within the WIM. The first ionization potential of O is 
quite close to that of H (13.595 eV and 13.614 eV, respectively) and the 
large $\mathrm{H}^{+} + \mathrm{O}^{0} \leftrightarrow \mathrm{H}^{0} + 
\mathrm{O}^{+}$ charge-exchange cross section keeps
$\mathrm{O}^{+}/\mathrm{O}$ nearly equal to $\mathrm{H}^{+}/\mathrm{H}$. 
Electron energies in $T_{e} \sim 10^{4}$~K gas are sufficient to excite 
the $\sim2$~eV ($^{3}P$---$^{1}D$) transition that results in the \oi\ 
$\lambda6300$ emission. Therefore, the intensity of this line relative to 
\ha\ is directly related to the amount of O$^{0}$, and thus H$^{0}$, 
relative to $\mathrm{H}^{+}$ in warm ionized gas \citep{Reynolds+98}.  In 
practice, nature conspires to make this observation very difficult, 
because \oi\ is also one 
of the brightest terrestrial emission lines in the night sky. 
Nevertheless, high-sensitivity, high-resolution spectroscopic measurements 
with WHAM have managed to resolve the Galactic emission from atmospheric 
emission to provide reliable measurements in a few select directions. 
These observations indicate that $\mathrm{H}^{+}/\mathrm{H}$ $> 90$\% for 
T $> 8000$~K \citep{Reynolds+98, Hausen+02-aj}.

The time scale for 
recombination at a typical WIM density of 0.1~cm$^{-3}$ is $\approx$ 
1~Myr.  This is shorter than the lifetimes of O~stars, the presumed 
ionizing sources, which implies that the photoionization rate of the 
neutral hydrogen atoms within the WIM is roughly balanced by the rate of 
hydrogen recombination.  In this case, the limit on 
$\mathrm{H}^{+}/\mathrm{H}$ implies an ionizing flux $> 10^5$ photons 
cm$^{-2}$ s$^{-1}$, consistent with the ionizing photon flux derived from 
the \ha\ surface brightness of \hi\ clouds.

Regarding heavier ions, observations reveal that in the WIM ions are 
generally in lower states of ionization than in classical O~star \hii\ 
regions \citep[e.g.,][]{MRH06}. The reason is not yet fully understood, 
since Lyman continuum photons emitted by massive O stars are almost 
certainly the primary source of ionization for the WIM (see \S 
\ref{sec:edgeon} and \S \ref{sec:faceon} below).  The lower ionization 
state could be due to a softening of portions of the spectrum as the 
radiation travels from the O stars to the WIM. Photoionization models 
\citep{WM04,HW03} show that the spectral processing of the radiation can 
be complex, with the radiation between the \hi\ and \hei\ ionization edges 
hardening with distance from the source, while the spectrum at higher 
energies softens.  Moreover, hot evolved low mass 
stars (white dwarfs) and interface radiation associated with the hot 
($10^{5-6}$~K) gas add harder photons to the mix (\S \ref{sec:interface}).  
Independently of the spectrum, the low ionization state of the WIM also 
could be the result of its low ionization parameter \citep{m86a}.

Constraints on the fluxes of higher energy (i.e., helium-ionizing) photons 
are from observations of the \hei\ recombination line at $\lambda5876$ and 
the \oiii\ $\lambda5007$ collisionally excited line. Both of these 
transitions are prominent in O~star \hii\ regions, where $I_{\nu}$($> 24$ 
eV) is high enough (and is \emph{known} to be high enough) to maintain 
He$^{+}$ (24.6 eV) and O$^{++}$ (35.1 eV) at appreciable levels. Even 
qualitatively, from the first attempts to detect these lines in the WIM 
\citep{RT95, Reynolds85b}, it was clear that these ions were not as 
abundant in the WIM. More recent WHAM observations found 
(\hei/\ha)$_{\mathrm{WIM}} \sim 0.5\ \times$ (\hei/\ha)$_{\rm{H~II}}$, 
which when combined with the fact that $\mathrm{H}^{+}/\mathrm{H}$ is near 
unity (see above), implies that He$^{+}$/He $ \alt 60\%$. The \oiii/\ha\ 
results are more varied, although the ratios are typically less ($\sim 
10\%$) those seen in \hii\ regions \citep{MadsenPhD, MRH06}.  The 
abundance of \oiii\ in other galaxies and in the interior regions of our 
Galaxy can be significantly higher than what is observed in the WIM near 
the sun \citep{Rand97, MR05}.

\subsection{Temperature} 
\label{sec:wimtemp}

The temperature of photoionized gas is set by a balance between heating 
and cooling.  Heat is injected by thermalization of the excess kinetic 
energy of the electron during the photoionization-recombination process 
\citep[see, e.g.,][]{o89}.  Other potential sources of heat could also be 
important, particularly at the low densities characteristic of the WIM 
\citep[e.g., see][]{RC92}. Cooling occurs primarily from the collisional 
excitation and subsequent radiative decay of metastable states (i.e., 
forbidden lines) of the trace ions \citep[see, e.g.,][for a detailed 
discussion]{o89}.  The detection of some of these ``cooling lines'' in 
combination with the H-recombination emission have been used to explore 
the temperature of the gas, as discussed in more detail below.  The 
observations have established that 1) on average the WIM is about 2000~K 
warmer than the denser, classical \hii\ regions and 2) there are 
significant variations in temperature within the WIM, most notably an 
increase in temperature with increasing distance away from the midplane, 
and more generally, with decreasing emission measure (or gas density). The 
reason for this temperature behavior of the WIM is not yet clear; it could 
indicate that photoionization is not the only important source of heat in 
the WIM \citep{RHT99} or that perhaps the spectrum of the ionizing 
radiation is modified as it propagates far from its source (see \S 
\ref{sec:clumpy}).

Although they vary in accuracy and difficulty, three tools are available to explore the temperature of the WIM through optical emission lines:

\begin{enumerate}

\item \textbf{\nii/\ha\ and \oii/\ha\ trace variations in $T_{e}$}:
N$^+$, O$^+$, and H$^+$ are the dominant states of ionization 
for these elements in the WIM. In addition, their emission lines have 
different dependences on temperature, so that changes in \nii/\ha\ and \oii/\ha\ 
closely track changes in $T_{e}$. Empirically, where $\iha < 1$ R, the 
brightnesses of the primary optical forbidden lines of \nii, \sii, and 
\oii\ become comparable to \ha, making these lines easier to detect and 
thus attractive diagnostic tools for exploring \emph{variations} in $T_{e}$.  
Calculating absolute temperatures is more uncertain due to necessary 
assumptions about the exact ionic fractions and elemental abundances.

\item \textbf{\nii\ $\lambda5755$/\nii\ $\lambda6583$ measures $T_{e}$ directly}:
The ratio 
of the ``auroral'' ($\lambda5755$) emission line (resulting from 
excitations to a metastable state 4.0 eV above ground) to that of the much 
brighter ``nebular'' ($\lambda6583$) transition (1.9 eV above ground) 
allows a derivation of $T_{e}$ with the fewest assumptions. Since this 
ratio involves the same ion, it is proportional to 
$e^{(\Delta E/k T_e)}$, where $\Delta E$ is the difference in the excitation energy of the two states. Near 8000~K, a 2000~K change in temperature 
produces about a factor of two change in the ratio.  However, because 
$I_{5755}/I_{6583} \sim 0.01$, this observation is extremely difficult for 
the WIM, where $\iha \sim 1$~R.

\item \textbf{Widths of resolved line profiles are proportional to $T_i$}:
Comparing the width of the \ha\ line to the width of an emission line from 
the heavier N$^+$ or S$^+$ ion can be used to separate the thermal from 
the nonthermal motions in the gas.  However, very high signal-to-noise 
ratio line profile measurements are needed because the derived value of 
the ion temperature $T_{i}$ is proportional to the difference of the 
squares of the line widths.

\end{enumerate}

Results using these techniques in recent observations are summarized below.

\subsubsection{\nii, \sii\, and \oii\ with respect to \ha}

Two robust statements can be made about the line ratio observations:

\begin{enumerate} 

%% Letters used here to enumerate to avoid confusion with section numbering

\item[(a)] \nii/\ha, \sii/\ha, and \oii/\ha\ increase with decreasing \iha. 
In the Galaxy, this rise is most dramatic below $\iha = 1$ R. The clearest 
examples of this are the large increases in the forbidden line 
intensities relative to \ha\ with increasing distance from the midplane, 
both in our Galaxy \citep{HRT99} and others \citep{Rand98, TDS00}.

\item[(b)] Values for \sii/\ha\ and \nii/\ha\ vary greatly, but are 
strongly 
correlated, often with a nearly constant \sii/\nii\ ratio over large 
regions. For the Galaxy, the ratio of \sii/\nii\ does not vary by more 
than about a 
factor of two, except in the vicinity of a discrete ionizing source 
(e.g., an O~star).

\end{enumerate}

As pointed out in \citet{HRT99}, this behavior suggests 
that changes in the line ratios are due primarily to changes in $T_{e}$.
Thus statement (b) above follows from the fact that the \sii\ and \nii\ 
lines have nearly identical excitation energies, so that

\begin{equation}
  \label{eq:siinii}
  \frac{I_{6716}}{I_{6583}} = 4.69\;
  \frac{\left(\frac{S^{\rule{0ex}{1ex}}}{H}\right)}
  {\left(\frac{N^{\rule{0ex}{1ex}}}{H}\right)}\,
  \frac{\left(\frac{S^+}{S}\right)}{\left(\frac{N^+}{N}\right)}\;
  e^{0.04/T_4}\;T_4^{-0.119},
\end{equation}

\noindent which is only a very weak function of $T_4\,(\equiv 
T_{e}/10^4$~K). From $T_{4} = 
0.5$ to $1.0$ with all else constant, \sii/\nii\ decreases only about 
11\%. This relationship also indicates that the relatively small but real 
variations of \sii/\nii\ that are observed in the WIM are tracing 
variations of $\mathrm{S}^{+}/\mathrm{N}^{+}$. Combined with the very 
different energies required for $\mathrm{S}^{+} \rightarrow 
\mathrm{S}^{++}$ (23.3 eV) and $\mathrm{N}^{+} \rightarrow 
\mathrm{N}^{++}$ (29.6 eV), we conclude that $\mathrm{S}^{+}/\mathrm{S}$, 
and especially $\mathrm{N}^{+}/\mathrm{N}$, vary little in the WIM and 
that the smaller (factor of two) variations in \sii/\nii\ are due 
primarily to variations in $\mathrm{S}^{+}/\mathrm{S}$. This is supported 
by photoionization models \citep[e.g.,][]{SHR2000}, which have shown that 
N$^+$/N $\approx 0.8$ over a wide range of input spectra and ionization 
parameters. 

On the other hand, the strong temperature dependence of the 
forbidden line intensities relative to \ha\ is illustrated by the 
relationship for the \nii/\ha

% Note: if line break is needed in this equation, recommend
% just before 'T_4' and add '\times' to end of first line.
\begin{equation}
  \label{eq:niiha}
  \frac{I_{6583}}{\iha} = 1.63\times10^5\;\left(\frac{H^+}{H}\right)^{-1}
  \left(\frac{N^{\rule{0ex}{1ex}}}{H}\right)
  \left(\frac{N^+}{N}\right)\;T_4^{0.426}\;e^{-2.18/T_4}.
\end{equation}

\noindent Because $\mathrm{N}^{+}/\mathrm{N}$ and 
$\mathrm{H}^{+}/\mathrm{H}$ vary little within the WIM, variations in 
\nii/\ha\ essentially trace variations in $T_{e}$.  Similar relationships 
can be written for \oii, \sii, and for other collisionally excited lines 
\citep[see][]{Otte+01}.

\begin{figure*}[tbp]
%%% RECOMMEND TWO-COLUMN WIDTH %%%
\begin{center}
\resizebox{\linewidth}{!}{\includegraphics{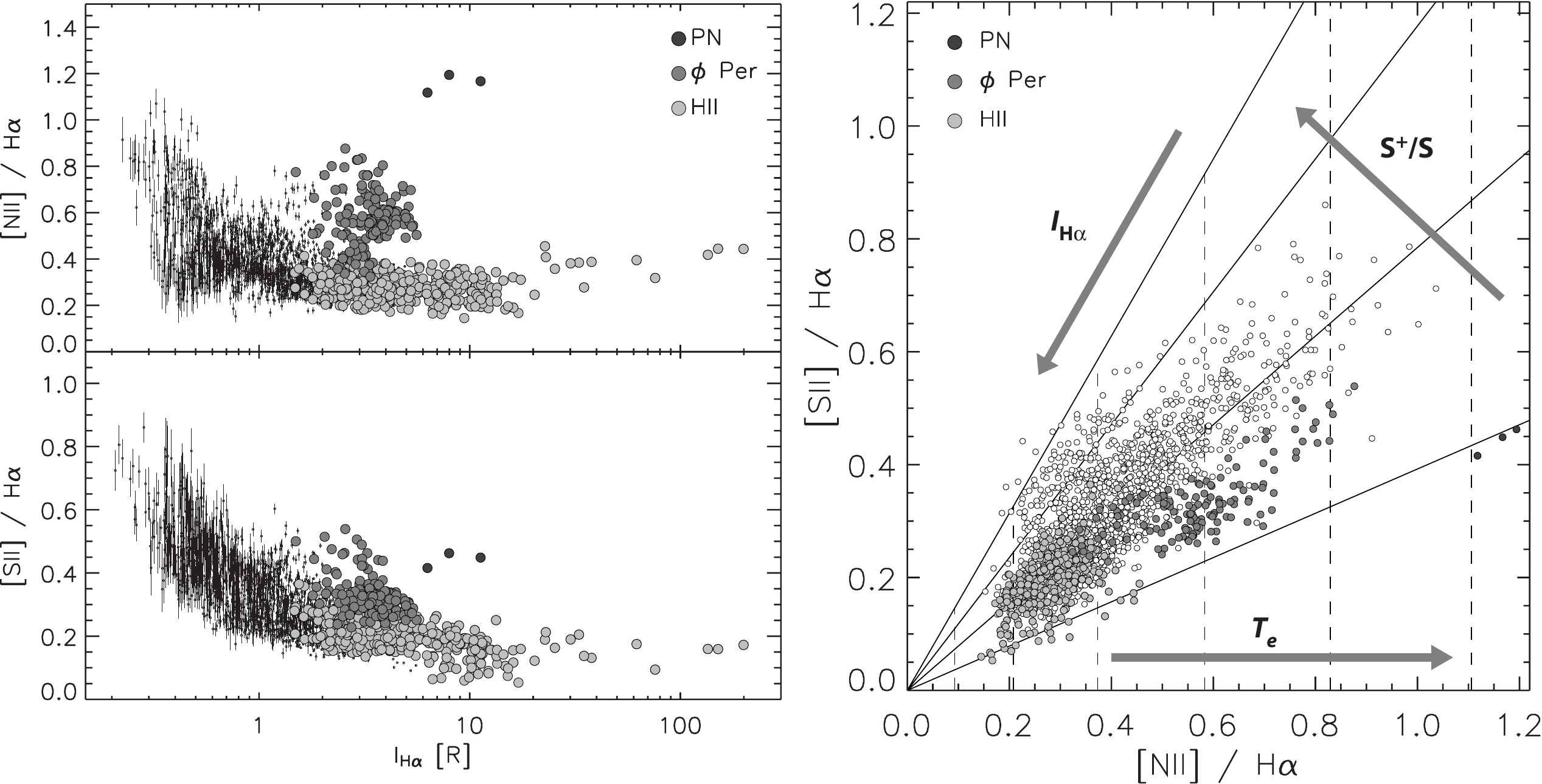}}
\caption{Diagnostic line ratio diagrams. A large portion of the Galaxy in 
the direction of the Perseus arm ($\ell = 130\degree$ to $160\degree$ and $b = 
-30\degree$ to $+30\degree$, approximately) has been surveyed in \ha, \sii, and 
\nii\ with WHAM. These panels show (\emph{left}) the relationship between the ratios 
versus the intensity of \ha\ as well as (\emph{right}) the relationship 
between the two ratios for the ``local'' gas component 
($|\vlsr| < 15$ \kms). A few specific spatial regions are highlighted with grayscale fill to 
show effects of local ionizing sources: the planetary nebula (PN) S216, 
the \hii\ region surrounding the B0.5+sdO system $\phi$ Per, and regions 
near O-star \hii\ regions. Other data points sample the WIM. In the diagram on the right, the vertical 
dashed lines represent $T_e$, 5000~K, 6000~K, 7000~K, 8000~K, 
9000~K, and 10~000~K, left to right.  The slanted solid lines 
represent S$^+$/S, 0.25, 0.50, 0.75, and 1.00, lowest to highest 
slope. These WIM data reveal significant variations in $T_e$ and 
S$^+$/S from one line of sight to the next. In contrast, classical 
\hii\ regions all cluster in the lower left corner of this diagram near 
\nii/\ha\ $\approx 0.25$, \sii/\ha\ $\approx 0.1$. From \citealp{MadsenPhD}.} 
\label{fig:greglr} 
\end{center}
\end{figure*}

Using Eq.\ 1 and 2, we can construct diagnostic diagrams as presented 
by \citet{HRT99} and \citet{MRH06} to estimate both $T_{e}$ and 
$\mathrm{S}^{+}/\mathrm{S}$ from observations of \nii/\ha\ and \sii/\ha, 
as shown in Fig.~\ref{fig:greglr} \citep[adapted from][]{MadsenPhD}. 
With sufficient velocity resolution it has even been possible to study 
variations in these parameters between different radial velocity 
components along the same line of sight. These results reveal that within 
the WIM there are variations in temperatures ranging between about 7000~K 
and 10~000~K and variations in $\mathrm{S}^{+}/\mathrm{S}$ between 0.3 to 
0.7. For comparison, the bright classical \hii\ regions all cluster near 
the lower left corner of the plot, \sii/\ha\ $\approx 0.1$ and \nii/\ha\ 
$\approx 0.25$, where $T_{e} = 6000$--7000~K and $\mathrm{S}^{+}/\mathrm{S} 
\approx 0.25$.

\oii\ at $\lambda3727$ has a larger excitation energy than \nii, making it 
even more sensitive to variations in $T_{e}$. Although inaccessible with 
WHAM, several extragalactic studies \citep{OGR02,Otte+01,TD00} have traced 
this line, and new instrumentation is starting to allow studies of \oii\ 
from the WIM  \citep{Mierkiewicz+06} of the Milky Way.  The \oii\ 
observations confirm that the line ratio variations are dominated by 
variations in $T_{e}$.

\subsubsection{\nii\ $\lambda5755$/\nii\ $\lambda6583$}

One of the most direct ways of measuring $T_{e}$ in ionized gas is to 
observe the ratio of two emission lines from the same ion but with very 
different excitation energies above ground. The $I_{4363}/I_{5007}$ ratio 
of \oiii\ in bright \hii\ regions is perhaps the most famous of these 
pairs. In the WIM, because the \oiii/\ha\ ratios are typically no more 
than 10\% of that in \hii\ regions, the isoelectronically similar \nii\ 
line 
ratio, $I_{5755}/I_{6583}$, has been used instead.

\begin{figure}[tbp]
\begin{center}
\resizebox{\linewidth}{!}{\includegraphics{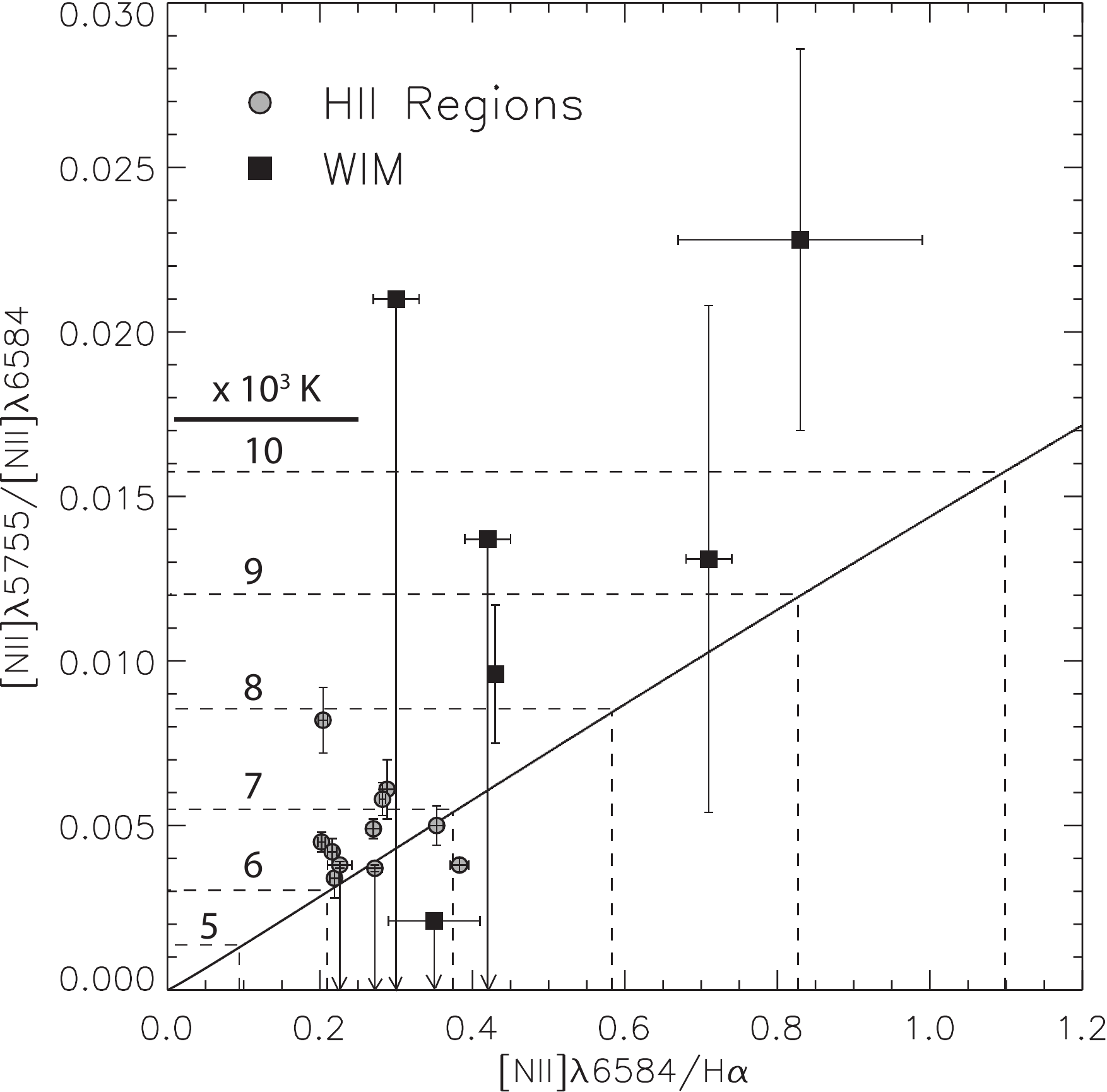}}
\caption{Elevated temperature in the WIM. Select directions toward 
 ($\Box$) brighter diffuse ionized regions show elevated line 
ratios in both \nii/\ha\ and \nii\ $\lambda5755/\lambda6584$ compared to 
 ($\bigcirc$) \hii\ regions. Dashed lines mark select temperatures 
derived from each line ratio while the diagonal line traces ``unity'' in 
this derived temperature space. WIM directions have higher derived 
temperatures using either line ratio, but are also systematically above 
the line of ``unity'', suggesting that measurable temperature 
inhomogeneities exist along these lines of sight. From 
\citealp{MadsenPhD}.}
\label{fig:greg5755}
\end{center}
\end{figure}

By detecting this extremely weak line, \citet{Reynolds+01a} and 
\citet{MadsenPhD} confirmed in select directions that $T_{e}$ in the WIM 
is indeed higher by about 2000~K than in the bright \hii\ regions. 
However, the details of the results reveal a more complicated temperature 
structure---perhaps not surprisingly. Although current measurements of the 
$\lambda5755$ line still have large uncertainties, 
Fig.~\ref{fig:greg5755} 
indicates that $T_{e}$ as inferred by the ratio of the \nii\ lines is 
systematically higher than that inferred by \nii/\ha\ in the same 
directions. This could be explained by temperature variations along the 
line of sight, since the $\lambda5755$ line (excitation energy 4 eV) 
would be 
produced preferentially in regions with higher $T_{e}$ compared to the red 
line (2 eV).

\subsubsection{Line widths}

If the intrinsic widths of emission lines can be measured accurately in 
ions having significantly different masses, then one can decompose the 
thermal motion (i.e., $T_{i}$) and non-thermal motion contributing to 
their widths. H 
and S are particularly good to use because they differ in mass by a factor 
of 32, resulting in a measurable difference in their widths. This method 
has been used with some success in both the WIM \citep{Reynolds85a} and 
\hii\ regions \citep{Reynolds88}.

\begin{figure}[tbp]
\begin{center}
\resizebox{\linewidth}{!}{\includegraphics{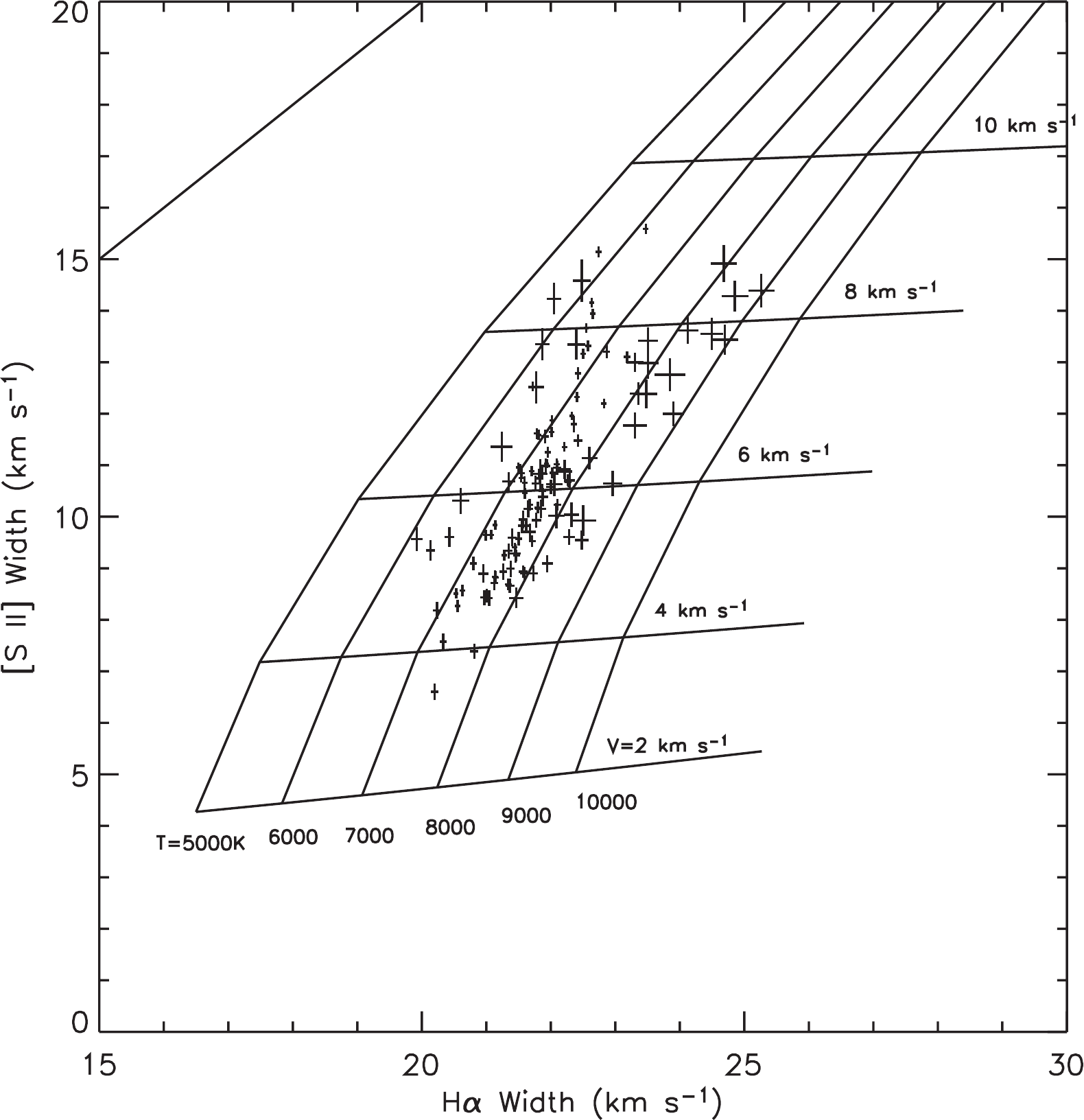}}
\caption{Temperature $T_{i}$ and the mode $v_{NT}$ (most probable value) 
of the non-thermal speed distribution of the diffuse \hii\ region 
surrounding $\zeta$ Oph. The primary axes show the measured widths 
and errors of \ha\ and \sii\ from 126 one-degree pointings obtained by WHAM within a 
12-degree diameter region around $\zeta$ Oph. A grid of temperatures and 
non-thermal velocities has been superimposed to show the distribution of 
$T_{i}$ and $v_{NT}$ within the \hii\ region. The apparent scatter 
in these
values is spatially correlated (see text). The line at the 
upper-left is the line of equal widths.}
\label{fig:zoph}
\end{center}
\end{figure}

The potential power of this technique is illustrated in 
Fig.~\ref{fig:zoph}, which shows line width data for the large, high 
Galactic latitude \hii\ region surrounding the O~star (O9.5V) $\zeta$~Oph. 
The \ha\ from this region spans an order of magnitude in intensity and at 
the fainter end becomes comparable to WIM emission at low Galactic 
latitudes ($\sim 10$ R). The $\zeta$~Oph \hii\ region is particularly good 
for line width studies because each emission line profile is very well 
described by a single Gaussian component. \citet{Baker+04} measured the 
widths of the \ha, \sii, and \nii\ emission lines from this region and 
derived accurate values for the temperature $T_{i}$ and the mode of the 
nonthermal speeds $v_{NT}$ as shown in Fig.~\ref{fig:zoph}. The large 
ranges in these parameters appears to be real, with a noticeable gradient 
of increasing $T_{i}$ (from 6000 to 8000~K) and decreasing $v_{NT}$ (from 
8 to 4 \kms) from the center to edge of the \hii\ region. The former is 
ascribed to the hardening of the radiation field with increasing distance 
from the source \citep[also, \S \ref{sec:clumpy} below]{WHR05}, while the 
latter could be explained by a slow (2 \kms) expansion of the \hii\ 
region. A future goal is to extend this method to the fainter and more 
kinematically complex WIM.

In summary, the evidence is now overwhelming from a variety of methods 
that in the WIM the temperature is elevated compared to the bright, 
classical \hii\ regions. In addition, the large variations in optical 
forbidden line strength relative to \ha\ within the WIM is dominated by 
changes in temperature rather than changes in ionic fractions or elemental 
abundances. When this is combined with the well-established result that 
the forbidden line ratios relative to \ha\ increase with decreasing \ha\ 
intensity (statement (a), above), we are led to the conclusion that the 
temperature increases with decreasing emission measure \citep{HRT99} and 
thus decreasing gas density. These temperature variations do not appear to 
be explained solely by photoionization heating of the gas \citep{WM04, 
RHT99}, suggesting additional heating sources for the WIM 
\citep{WD01,MS97,RC92} that begin to dominate over photoionization heating 
at low densities ($\sim 10^{-1}$ cm$^{-3}$).

\subsection{Warm ionized and neutral gas}
\label{hihii}

The fact that hydrogen is nearly fully ionized within the \ha\ emitting 
gas (\S \ref{sec:ionization}) implies that $\mathrm{H}^{+}$ and 
$\mathrm{H}^{0}$ are primarily confined to separate regions. With the 
advent of velocity resolved \ha\ surveys (i.e., WHAM), it is now possible 
to begin to explore the relationship between the diffuse ionized gas and 
neutral gas in the interstellar medium. Is the WIM the ionized portion of 
a low density ``intercloud medium'' \citep{MC93} or is it mostly confined 
to the surfaces of neutral clouds, the transition region between cooler 
gas and a much hotter ``coronal'' temperature medium \citep{MO77}?  Do the 
ionized and neutral phases cycle from one to the other?  \citet{CH03} and 
\citet{Lockman04}, for example, have suggested that portions of the 
ionized medium may condense into neutral clouds. However, to date there 
have been no observational studies of the $\mathrm{H}^{+}$ -- 
$\mathrm{H}^{0}$ connection. In a general qualitative sense, a comparison 
of the optical line profiles with the radio 21~cm (\hi) profiles (e.g., 
\citealp{LDS}), indicates that at high latitudes the ionized gas tends to 
be correlated in space and velocity with the so-called warm neutral medium 
(WNM), the wide-spread, $T \sim 10^3$~K phase of the \hi\ generally 
associated with broad 21~cm profiles.  There is very little correspondence 
between the \ha\ and the narrow-line 21 cm emission components associated 
with the colder ($T \sim 10^2$~K), denser \hi\ clouds.  In regions of the 
sky that contain anomalous velocity structures, specifically, the 
intermediate and high velocity clouds that are not co-rotating with the 
Galactic disk, the correlation is quite strong \citep{HRT01, TRH98}. The 
optical and radio emission components are centered at nearly the same 
velocities (within roughly 5 \kms) and have comparable velocity extents in 
cases of complicated, blended profiles. Few regions seem to contain only a 
WNM or only a WIM component. This relationship has been hinted at through 
various detailed absorption line and emission line studies over the last 
decade \citep{SF93, HSS03, RTK95}, which together with observations of the 
\oi\ $\lambda6300$/\ha\ line intensity ratio (\S \ref{sec:wimtemp} above)  
suggest that a significant amount of the $\mathrm{H}^{+}$ is associated 
with nearly fully ionized regions in contact with (or at least adjacent 
to) regions of warm primarily neutral hydrogen.  Hopefully a systematic 
examination of the kinematic and spatial correlation between these ionized 
and neutral phases of the medium will be carried out soon.

Although there is a close correspondence on the sky and in velocity 
between the warm neutral and ionized emission lines, their intensities do 
not appear to be correlated. This has been examined in detail only toward 
two intermediate velocity \hi\ clouds, Complexes L and K 
\citep{Haffner05,HRT01}. In both cases, the 
column density of the neutral hydrogen \nhi\ and the \ha\ intensity, \iha, 
are uncorrelated. Whether this holds for more local gas in the Galactic 
disk still needs to be fully explored.  A straightforward explanation for 
this lack of correlation is the fact that the intensity of the \ha\ is 
determined solely by the flux of ionizing radiation incident on the warm 
\hi\ cloud \citep{RTK95}, which of course, is independent of the cloud's 
column density because the \hi\ clouds are optically thick to the Lyman 
continuum photons.

\subsection{The role of superbubbles}
\label{bubbles}

One of the basic questions concerning the nature of the WIM is how 
ionizing photons from the O stars can travel hundreds of parsecs through 
the disk and into the halo.  A fractal morphology of the interstellar 
medium is one possibility (see \S \ref{sec:clumpy}).  Another is the 
existence of enormous, \hi-free bubbles surrounding some of the O stars, 
which allow the Lyman continuum photons to travel through the cavity to 
ionize its distant walls \citep[e.g.,][]{RO79, NI89, MFP06, PLS07}.  A 
WHAM study of one of these bubbles, the Perseus superbubble \citep{MRH06, 
Reynolds+01b}, has shown that a luminous O-star cluster near the midplane 
can indeed produce wide-spread, nearly WIM-like ionization conditions out 
to distances of 1000~pc or more from the ionizing stars.  However, the 
\nii/\ha\ and \sii/\ha\ ratios of the superbubble wall are not quite as 
large as the ratios observed in the surrounding WIM, suggesting that 
bubble size, gas density within the shell, supplemental heating, and/or 
the flux and spectrum of the radiation escaping the O-star cluster may 
also be important in setting the conditions of the ionized gas.

\section{Views from the outside: diffuse ionized gas and star formation 
rates in edge-on galaxies}
\label{sec:edgeon}

In galaxies other than our own, the wide-spread $\mathrm{H}^{+}$ is most 
often referred to as the diffuse ionized gas, or DIG. While its physical 
properties can be measured in much more detail for the interstellar medium 
of our Galaxy, the detection of the DIG in other galaxies provides the 
much needed ``outside'' perspective. \citet{Rand97, Rand98}, for example, 
was able to detect the DIG far into the halo of the edge-on spiral galaxy 
NGC\,891 and found that not only was \sii/\ha\ anomalously high, as 
observed in the solar neighborhood, but that \sii/\ha, \nii/\ha, and even 
\oiii/\ha\ increased significantly with increasing distance from the 
midplane.  This opened up a broader discussion about the ionization and 
heating processes within the gas and their variation with location within 
a galaxy.

Furthermore, observations of other galaxies provide important new 
information about the links between the DIG and global properties of 
galaxies, such as their star formation activity. In particular, the 
structure of this gas perpendicular to the galactic plane, that is, in the 
main direction of the gravitational potential, is an excellent tracer of 
the dynamics of the interstellar medium driven by energetic galactic 
processes.  The observations corroborate the picture of a dynamic 
interstellar medium driven by multiple and clustered supernova, producing 
the so-called disk-halo interaction between the star formation regions 
near the midplane and energized interstellar matter that extends a 
kiloparsec or more above the disk. Early discussions of these ideas can be 
found in \emph{Supernovae, the Interstellar Medium, and the Gaseous Halo: 
The Swashbuckler's Approach} by \citet{Heiles86}, \emph{The disk-halo 
interaction - Superbubbles and the structure of the interstellar medium} 
by \citet{NI89}, and \emph{Galactic worms} by \citet{KHR92}.

\begin{figure}[tbp]
\resizebox{\linewidth}{!}{\includegraphics{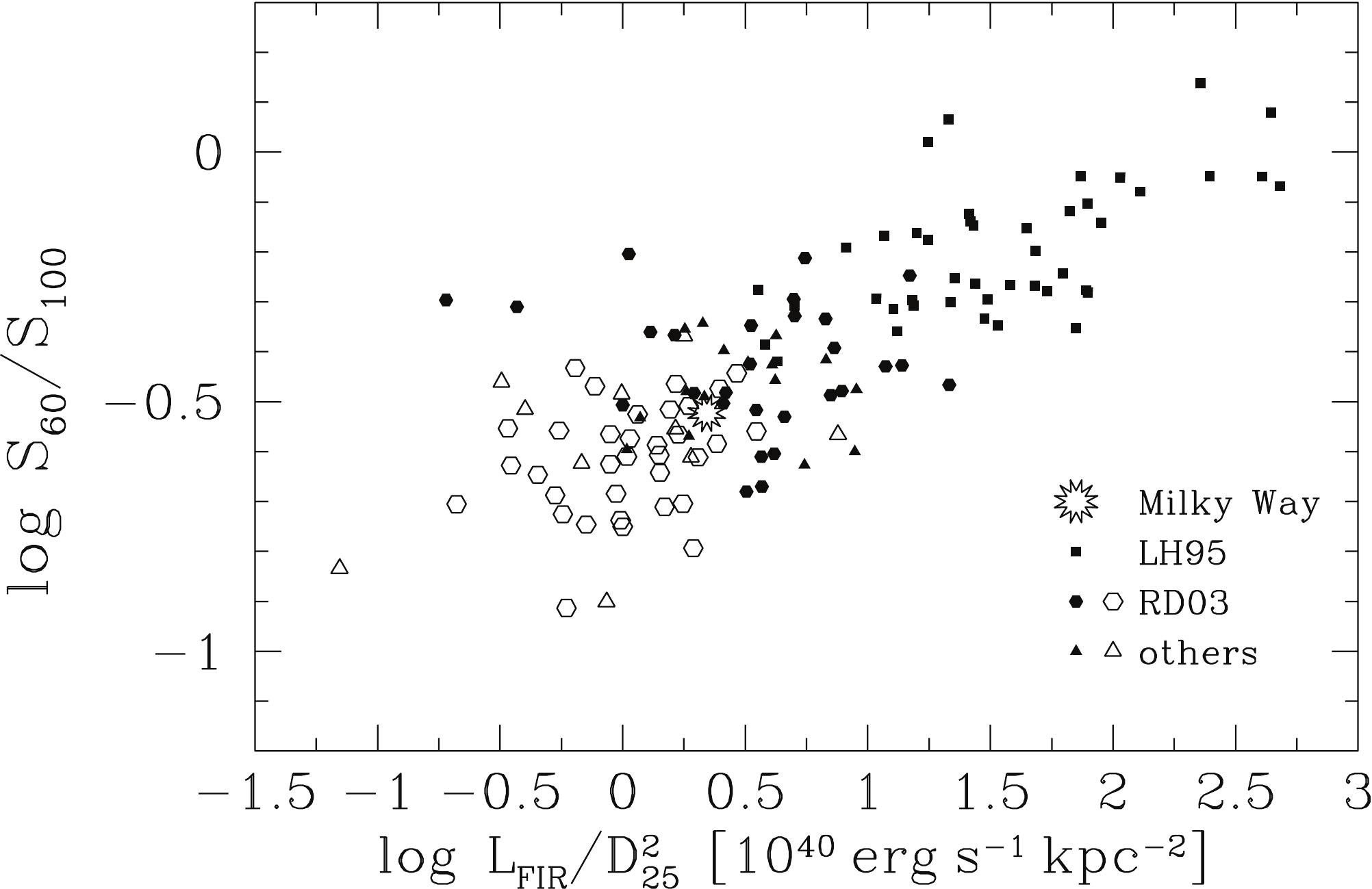}}
\caption{The dependence of DIG detections in halos of galaxies on the normalized star formation rate (per unit area). The ratio of the FIR fluxes at 60$\mu$m and 100$\mu$m ($S_{60}/S_{100}$) is given versus the FIR luminosity normalized to the optical galaxy diameter at 25$^{\mathrm{th}}$ \,mag in units of $10^{40}\rm{\,erg\,s^{-1}\,kpc^{-2}}$. Filled symbols indicate detections of $\mathrm{H}^{+}$ gas in the halo while open symbols represent non-detections. Galaxies from the ($\blacksquare$) starburst sample of \citealp{LH95} are included with more normal spirals. Integrated values for the Galaxy are taken from \citet{BDT90} and \citet{CM89} for the FIR. A Galactic disk radius of 22.5\,kpc is assumed. Adopted from \citealp{RD03a}.
\label{fig:dddplusmw9}
}
\end{figure}

\begin{figure}[tbp]
\resizebox{\linewidth}{!}{\includegraphics{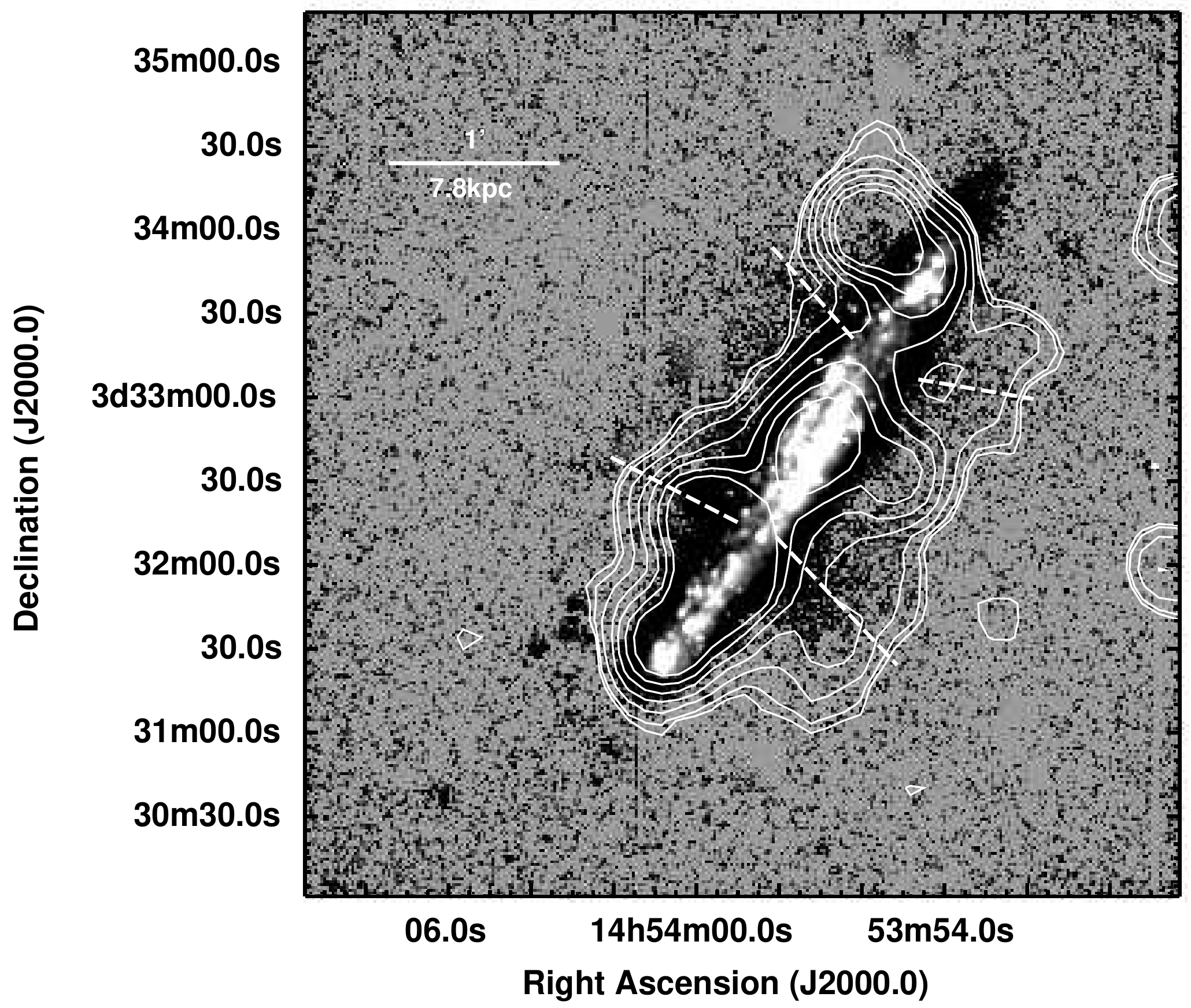}}
\caption{The X-ray halo NGC\,5775 from integrated EPIC 
XMM-Newton observations exhibits
    substructure correlating on galactic scales with filaments
    of  diffuse ionized gas and spurs in the radiocontinuum.  Adopted from \citealp{TPR06}.}
\label{fig:ngc5775}
\end{figure}

\subsection{Diffuse ionized gas in the halo and star formation in the 
disk}
\label{starform}

From an observational point of view, the study of the warm 
$\mathrm{H}^{+}$ in galactic halos---the extraplanar diffuse ionized
gas---is a good start toward understanding the nature of gaseous halos 
and the disk-halo connection.  Of all tracers of halo gas, the warm 
ionized gas is the easiest to observe with regard to sensitivity and 
resolution \citep{Dettmar92, Dettmar98, Rand97, Rand98}, allowing us to 
study the global influence of the energy released by young and massive 
stars into the interstellar medium.

A halo component of the DIG for external galaxies was first discovered in 
NGC\,891 \citep{Dettmar90, RKH90}, followed by studies of a number of 
other galaxies \citep[e.g.,][]{PBS94, Rand96, RD00}. In \citet{RD03a, 
RD03b} a large sample of edge-on spiral galaxies clearly demonstrates a 
correlation of DIG in the halo with the star formation rate (SFR) in the 
disk \citep[see also][for comparison]{MV03}. This survey covered a broad 
range in SFR extending the observations to less active galaxies. Until 
then, emphasis had been given mainly to galaxies with high SFR or even 
starburst galaxies \citep{LH95}. This relationship between the presence of 
DIG and far infrared (FIR) emission (a measure of the SFR) is shown in 
Fig.~\ref{fig:dddplusmw9}. If the star formation rate per unit area is 
low, as determined by the FIR luminosity normalized to the disk surface 
area, the presence of halo DIG indeed is observed to diminish.

The FIR luminosity per unit area therefore seems to be a promising 
indicator for the presence of halo $\mathrm{H}^{+}$ and suggests that 
there is a minimum SFR per area required to drive the disk-halo 
interaction \citep{RD03a}. Given the known large uncertainties 
in the normalization of supernova rates from FIR-fluxes, a reasonable 
estimate for the local break-out condition is on the order of $\sim$15 SNe 
over the lifetime of the O~star association. An obvious shortcoming of 
this analysis is the use of global measurements for the FIR luminosities. 
Future observations with higher angular resolution, for example, with the 
Spitzer-satellite, could allow a much better comparison of the local SFR 
with associated halo gas properties.

\subsection{The disk-halo connection and hot gas}
\label{hothalos}

The scenario of the disk-halo connection predicts outflows of hot 
($10^6$~K), supernova-created gas away from the disk and into the halo via 
superbubbles (or ``chimneys'').  In this way, transparent (i.e., 
\hi -free) pathways are then also provided for the 
transport of ionizing radiation into the halo and beyond.  For
galaxies with strong starburst activity this is well supported by 
observations \citep[see, e.g.,][]{CBV02, SHC04a, SHC04b, VCB05} 
and described well by theory and numerical modeling \citep[e.g.,][]{MF99}. 
However, for more normal (i.e., non-starburst) galaxies, we 
have little direct evidence for such outflows.  High resolution 
ground-based \citep[e.g.,][]{HS00} and HST imaging 
\citep[e.g.,][]{RDW04} studies of the DIG distribution in edge-on galaxies 
does not reveal the number and specific morphology of superbubble- or 
chimney--like structures predicted by some models \citep{NI89}.  Such 
structures have been found, but their number 
is small. The difficulty in finding them could be an observational 
problem, e.g., confusion due to superpositions along the line of sight, or 
it could mean the the current models of the disk-halo connection need 
further work. Thus, the idea of actual mass exchange between disk and halo 
is still a working hypothesis only.

\begin{figure}[tbp]
\begin{center}
\resizebox{\linewidth}{!}{\includegraphics{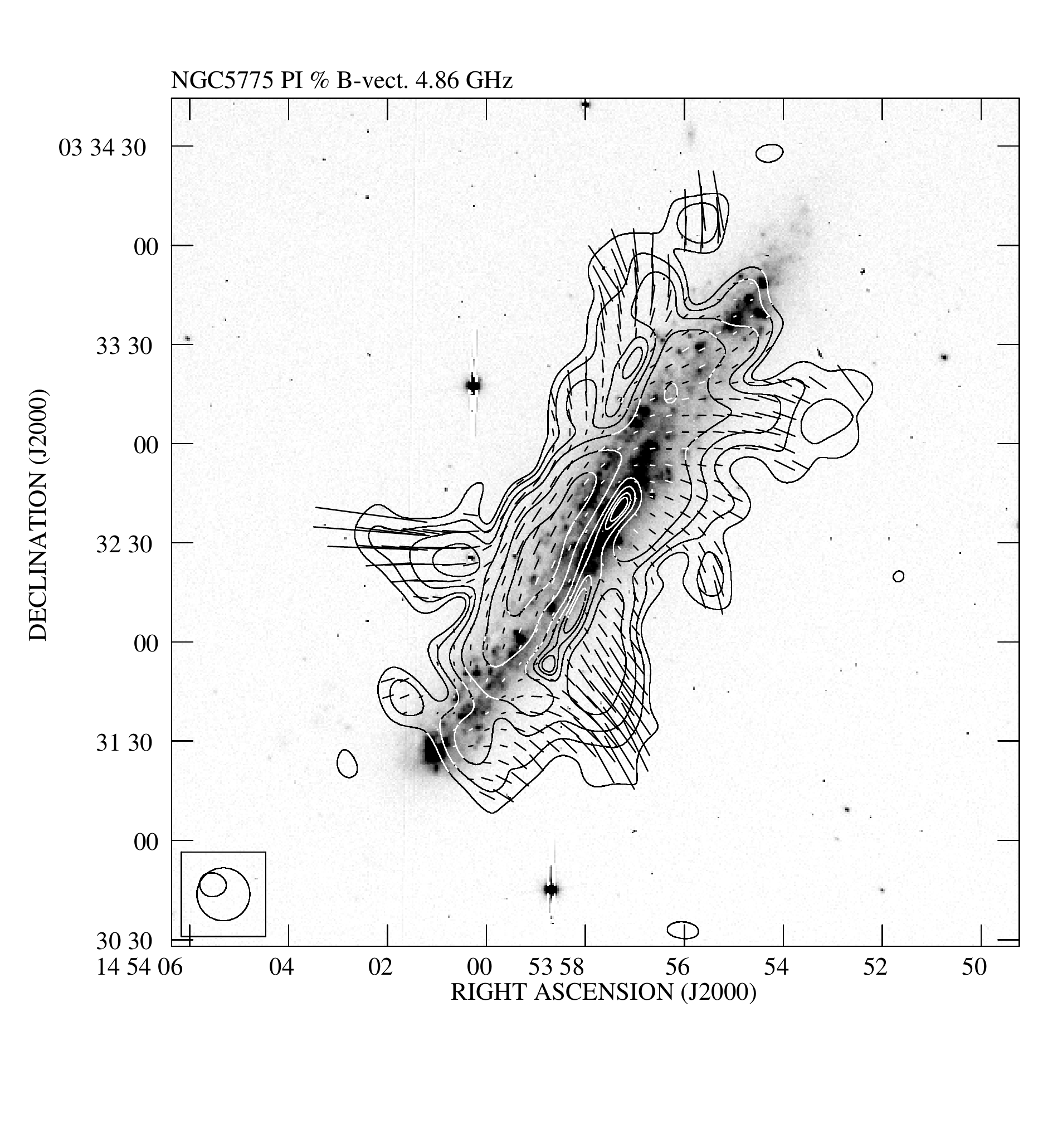}}
\caption{VLA radiocontinuum map of NGC\,5775 at 6cm. The polarized intensity
  is given in contours on a grey scale representation of the \ha\ 
emission. The 
bars represent the magnetic field direction and strength. Adopted 
from \citealp{TDS00}, courtesy M. Soida.} 
\label{fig:pi5775}
\end{center}
\end{figure}

On the other hand, there is circumstantial evidence for hot gas outflows 
in more normal disk galaxies that is indicated by the detection of diffuse 
interstellar X-ray emission, both in our Galaxy \citep[e.g.,][]{SEA07} 
and others \citep[e.g.,][]{WIW01}.  In a sample by 
\citet{RCS03} that 
included disk galaxies with more normal star formation rates, 
a correlation was found between the galaxy's global X-ray emission, 
presumably associated with the very hot ($10^6$~K) gas located inside 
supernova remnants and superbubbles, and the amount of star formation 
activity in the disk. With the current generation of X-ray satellites it 
is now possible to extend the X-ray studies from the starbursting and 
luminous galaxies to less active galaxies.  One result of a small survey 
with XMM-Newton is shown in Fig.~\ref{fig:ngc5775}. The EPIC X-ray image 
of the galaxy NGC\,5775 is integrated over all energies and summed over 
all detectors. It demonstrates the presence of smaller scale X-ray 
emission features associated with \ha\ filaments far above the disk.

The diffuse $\mathrm{H}^{+}$ in galactic halos is, in most of the observed 
cases, also associated with synchrotron radiocontinuum halos, suggesting 
that there is also a relationship between DIG and the presence of cosmic 
rays and magnetic fields in the halo.  For example, NGC\,5775 shows a 
prominent radio continuum halo, and a study of its polarization 
characteristics with the Very Large Array (VLA) radio synthesis telescope 
reveals a highly structured large-scale magnetic field with an 
ordered 
component to the field lines which opens up into the halo 
(Fig.~\ref{fig:pi5775}). This particular structure of the magnetic field 
could be of interest for the interpretation of global dynamo theory in 
galaxies and is discussed in more detail by \citet{BBM96}. The condition 
that allows cosmic rays to break out of the disk as a function of star 
formation rate per unit area is discussed by \citet{DLG95}. A detailed 
study of the kinematics of the DIG in NGC\,5775 is presented by 
\citet{HRB06}.  A compilation of observations of edge-on galaxies in 
H$\alpha$, radio continuum, and X-rays is given in \citet{Dettmar98} and 
\citet{RD03a, RD03b}.

In summary, studies of edge-on galaxies have shown that warm ionized gas 
halos are found in galaxies where the SFR per unit area is sufficiently 
high. Typically these layers of extraplanar DIG can be traced out to 
distances of z $\approx$ 1000--2000\,pc, sometimes even up to 5000\,pc or 
more from the mid-plane of the disk. The $\mathrm{H}^{+}$ halos seem to 
be associated with halos of cosmic rays and X-ray emitting plasma, which 
is expected if superbubble or chimneys provide absorption free pathways 
for hydrogen ionizing photons as well as being conduits for the vertical 
transport of hot gas and cosmic rays away from the disk and into the halo.

\section{Views from the outside: The source of the diffuse ionized gas in 
face-on galaxies}
\label{sec:faceon} 

Observations of edge-on and face-on galaxies clearly provide complementary 
perspectives of the distribution of the DIG. Specifically, face-on 
galaxies show the surface brightness of emission lines from the DIG across 
the face of the galaxy, which makes it possible to explore relationships 
(if any) between the properties of this plasma and the locations and 
ionizing fluxes of the hot, massive O stars, which are the most powerful 
ionizing agents in disk galaxies and are the presumed ionizing source for 
the DIG. In general, the observations reveal a strong relationship, with 
the \ha\ flux from the DIG comparable to that from the classical O~star 
\hii\ regions in the galaxy.

The presence of diffuse interstellar \ha\ emission in face-on spirals was 
first noted by \citet{Monnet71}, who derived a temperature of 7000~K, an 
emission measure of about 35 cm$^{-6}$ pc, and a density near 0.5 
cm$^{-3}$ for the emitting gas.  Modern detector technology 
(i.e., CCDs) has pushed the detection of diffuse $\mathrm{H}^{+}$ to 
fainter regions and has allowed the study of other emission lines, which 
has provided insight into the relationship between the diffuse ionization 
and the O stars.

\subsection{Radiation from O~stars and the surface brightness of the DIG}
\label{ferguson}

\begin{figure}[tbp]
\begin{center}
\resizebox{\linewidth}{!}{\includegraphics{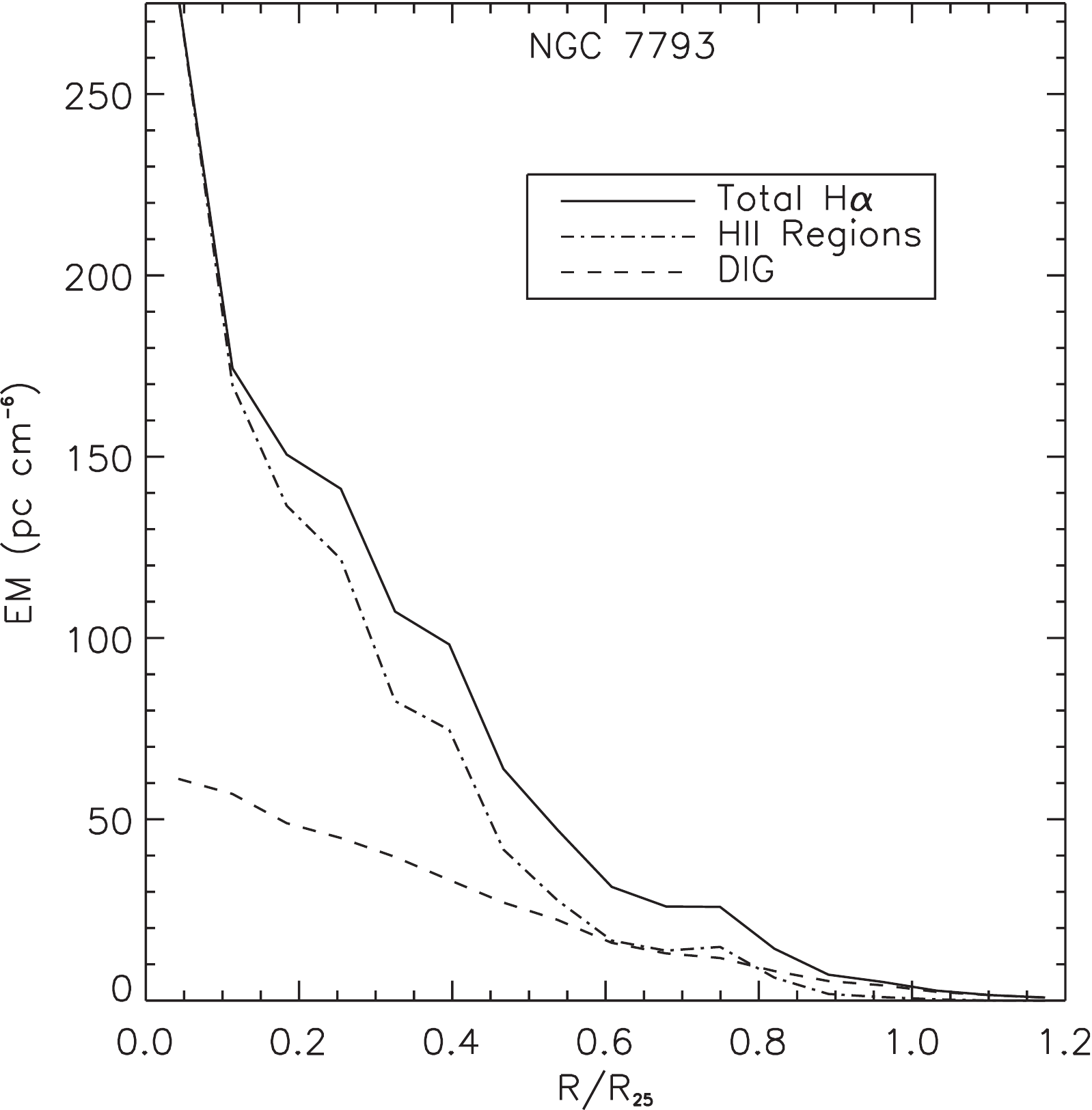}}
\caption{Deprojected profiles of the (\emph{solid}) total \ha\ surface brightness
  into (\emph{dashed-dotted})  \hii\ region and (\emph{dashed}) DIG emission
  for NGC\,7793. From \citealp{FWG96}.
}
\label{fig:beckman1}
\end{center}
\end{figure}

Although it was not understood how Lyman continuum photons could have free 
paths of hundreds of parsecs and more in galaxies with typical 
interstellar \hi\ densities of $\sim$1 cm$^{-3}$, ever since the discovery 
of diffuse $\mathrm{H}^{+}$, O~stars have been considered the prime 
candidate for the ionization.  Other known energy sources simply fall 
short in total power \citep[e.g.,][]{Reynolds84}.  A key observational 
step that 
connected the diffuse ionized gas to radiation from O~stars was carried 
out by \citet{FWG96}, who showed a quantitative relationship between the 
DIG and the surface brightness distribution of the bright, O~star \hii\ 
regions across a galactic disk. An example of their work is given in 
Fig.~\ref{fig:beckman1} (their Fig.~4a), where it is clear that the mean 
radial surface brightness profile of the DIG in \ha\ emission tracks that 
of the \hii\ regions. Their study included a careful, quantitative 
comparison of the energetic requirements for the \ha\ emission in the DIG 
with the ionizing radiation and the mechanical energy inferred to be 
emitted by the O~stars in the \hii\ regions of the two galaxies measured. 
They found that the mechanical energy clearly fell short, by more than a 
factor of three. They also found that the maximum contribution of local 
sources of ionizing radiation in the DIG from stars cooler than spectral 
type O8 also fell short of the luminosity required, but by a smaller 
factor. They concluded that the O~star populations of the clusters 
producing the \hii\ regions are the likeliest main source of the radiation 
that ionizes the DIG.

\begin{figure*}[tbp]
%%% RECOMMEND TWO-COLUMN WIDTH (resulting in a full page) %%%
\begin{center}
\resizebox{\linewidth}{!}{\includegraphics{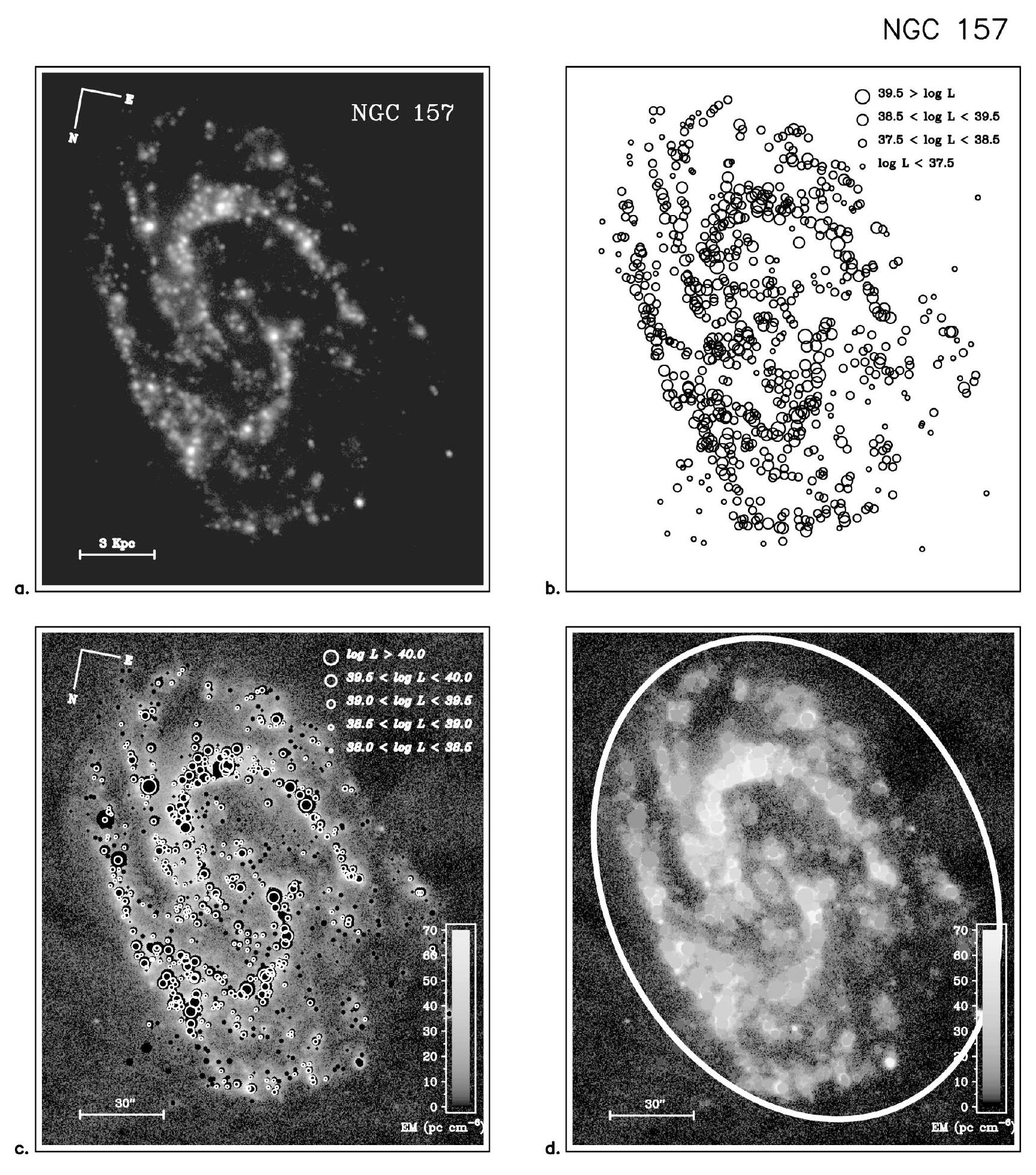}}
\caption{Steps in the quantification of the total \ha\ luminosity
   from the DIG for the representative disk galaxy NGC\,157.
   (\emph{a}) Continuum subtracted \ha\ image. (\emph{b}) Schematic form of \hii\
   region catalog, giving position and key to the \ha\ luminosity
   of each region. (\emph{c}) Diffuse \ha\ map after subtracting off the
   catalogued \hii\ regions. The brightest \hii\
   regions indicated by circles. (\emph{d}) Measurement of upper limit for
   DIG. \hii\ regions are blanked off, then each is assigned a local
   value of DIG surface brightness. Ellipse shows limit of integrated
   DIG flux measured. From \citealp{ZRB00}.
}
\label{fig:beckman2}
\end{center}
\end{figure*}

In two papers dedicated to testing the hypothesis that escaping
Lyman continuum photons from the classical \hii\ regions surrounding 
O~stars can be sufficient to ionize the DIG, \citet{ZRB00, ZBR02} took 
another important step forward by identifying and
   classifying the \hii\ regions in a set of photometric maps of disk
   galaxies in \ha. This process is illustrated in 
   Fig.~\ref{fig:beckman2}, taken from
   \citet{ZRB00}, which shows in the upper
left, the original continuum--subtracted \ha\ image of the nearly face-on 
spiral galaxy NGC\,157, followed clockwise by a
   schematic representation of the positions and luminosities of the 
   classified \hii\ regions, a surface brightness map of the DIG in \ha, 
   with the \hii\ regions subtracted off, and finally a map used for 
   quantifying the DIG. In the case of this last frame, the DIG is 
   measured by integrating the \ha\ surface brightness over the full disk, 
   using the values outside the \hii\ region boundaries and local mean 
   values inside each \hii\ region.  Some key points in the method for 
   deriving these maps and quantifying the DIG emission include the 
   following. First a catalog of \hii\ regions was prepared, using a 
   semiautomatic, but interactive method to measure their \ha\ 
   luminosities, effective radii and central positions \citep{RZH99}, down 
to a lower limiting luminosity. These \hii\ regions were 
   delimited and their emission subtracted from the total image by masking 
   those pixels occupied by the \hii\ regions, but with a final refinement 
   that contiguous pixels with surface brightness higher than a set limit 
   are also subtracted off.

\begin{figure}[tbp]
\begin{center}
\resizebox{\linewidth}{!}{\includegraphics{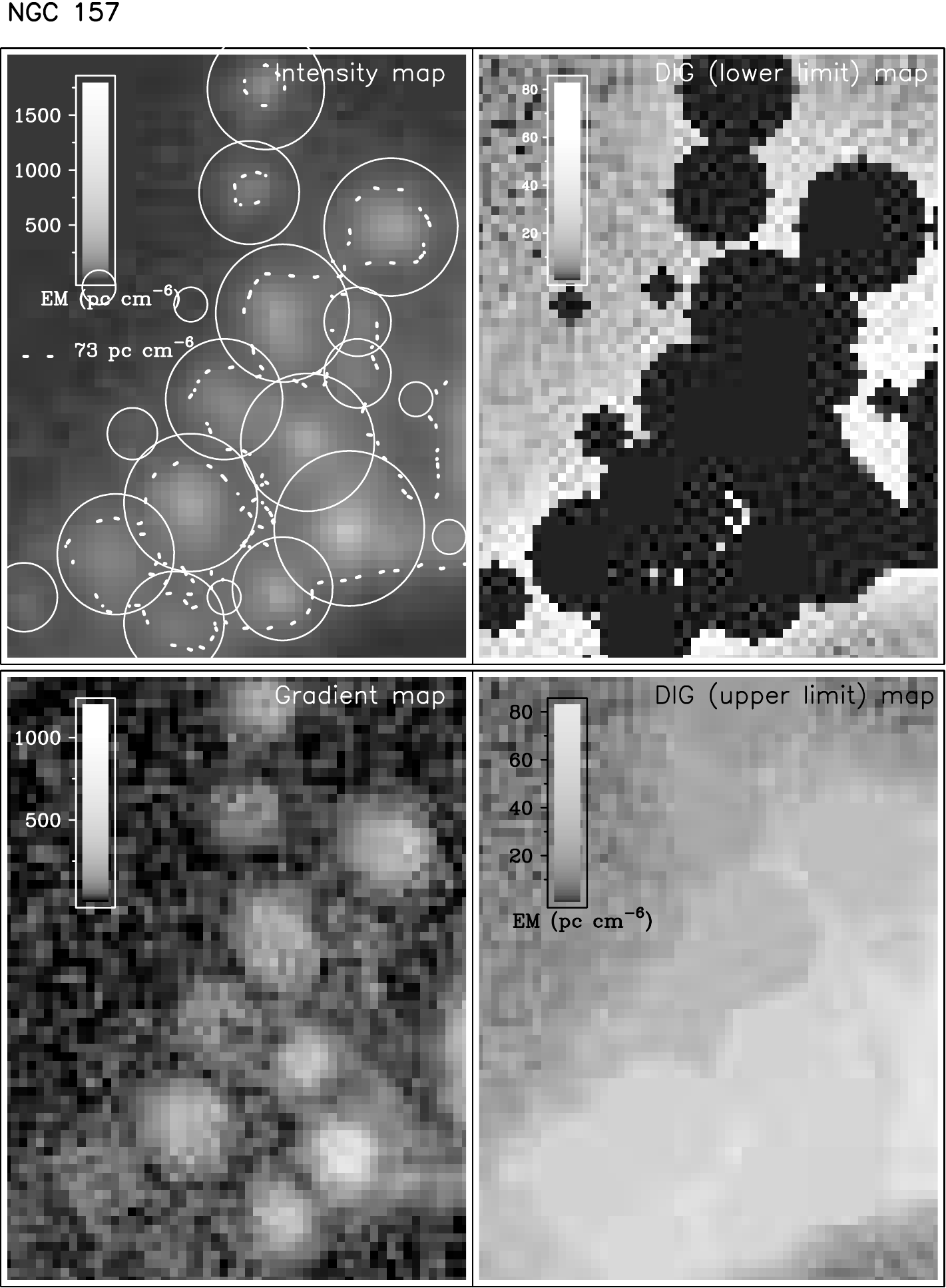}}
\caption{Technique used for separating \hii\ region emission from DIG
   emission. (\emph{a}) Portion of \ha\ image of NGC\,157 with crowded 
field
   (on a spiral arm). Grey scale in emission measure. Circles show
   mean catalogued radii of \hii\ regions. Dashed line at 73~pc~cm$^{-6}$,
   cut--off applied to avoid contamination of DIG by \hii\ regions.
  (\emph{b}) Diffuse emission after subtracting off \hii\ regions and applying
   surface brightness cut--off. (\emph{c}) Same portion of image in units    of
   \ha\ surface brightness gradient. Applying uniform cut--off at
   12.4 pc~cm$^{-3}$ pixel$^{-1}$ (0.28''/pix) yields \hii\ region boundaries equal to those
   found in (a), which confirms this separation technique. (\emph{d}) Map as
   in (b), but with \hii\ region mask filled at level of local DIG, giving
   upper limit case for total galaxy DIG luminosity (see text). From
   \citealp{ZRB00}.
}
\label{fig:beckman3}
\end{center}
\end{figure}

\begin{figure}[tbp]
\begin{center}
\resizebox{\linewidth}{!}{\includegraphics{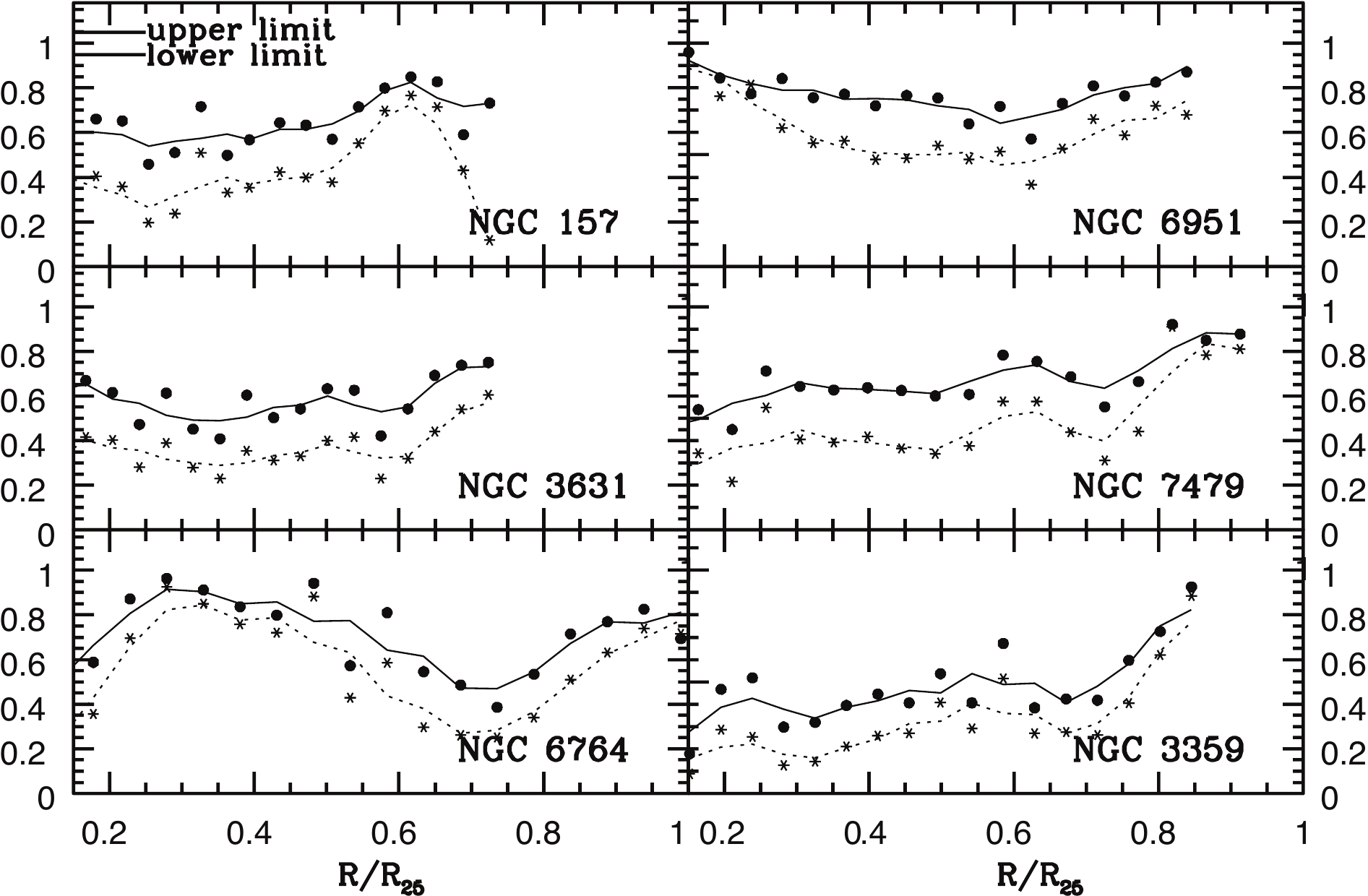}}
\caption{Radial variation of the ratio of integrated DIG luminosity in
    \ha\ to the total luminosity for six spiral galaxies (within
    0.15R/R$_{25}$ crowding precludes accurate estimates). A canonical value
    for the integrated ratio of $\ge$50\% is found, with a total fractional
    area subtended by the DIG of $\sim$80\%. From \citealp{ZRB00}.
}
\label{fig:beckman4}
\end{center}
\end{figure}

This procedure is illustrated in Fig.~\ref{fig:beckman3} \citep{ZRB00}, 
   which
   shows how this criterion for separating \hii\ region from DIG emission
   coincides very well with an alternative criterion in which the 
   boundaries
   of an \hii\ region are defined by a limiting value of surface brightness
   gradient. Having separated the \hii\ regions, one can then define the
   total DIG luminosity in one of three ways: (a) integrating the remaining
   surface luminosity after applying the \hii\ region mask; (b), as in (a)
   but then adding a contribution from the areas of the \hii\ regions,
   assuming this is  
   proportional to their projected areas times their local DIG surface
   brightness, or (c), as in (a) but making the contribution proportional
   to the area of the \hii\ regions times the mean DIG surface brightness
   outside the regions 
   across the disk of the galaxy.
   Modes (a), (b), and (c) give respectively lower
   limits, upper limits, and approximate estimates for the total DIG
   emission from the observed galaxies. In Fig.~\ref{fig:beckman4}, we 
   show the results 
   of
   this method of estimating the DIG, for six galaxies, shown as the ratio
   of the DIG \ha\ luminosity to the total luminosity for the galaxy
   plotted in terms of galactocentric radius. 

We can see that the DIG
   emits around a half of the total \ha\ output, that there are
   systematic modulations of this tendency with radius, and that there is
   slight tendency for the fraction to increase with radius. We also note
   that the projected area of the disk occupied by the DIG is of
   order 80($\pm$10)\% for all the objects shown in 
   Fig.~\ref{fig:beckman4}. In a separate study of over 100 
   galaxies, 
   \citet{Oeyetal07} found that the amount of DIG \ha\ to total \ha\ from
   a galaxy ranged from 20\% to nearly 100\% with a mean near 60\%.  
\citet{Voges06} has presented results suggesting an inverse correlation 
between the DIG \ha\ fraction and the star formation rate per unit area.

\subsection{An escape model for Lyman continuum propagation}
\label{lcpropagation}
      
\begin{figure*}[tbp]
%%% RECOMMEND TWO-COLUMN WIDTH (to appreciate figure detail) %%%
\begin{center}
\resizebox{\linewidth}{!}{\includegraphics{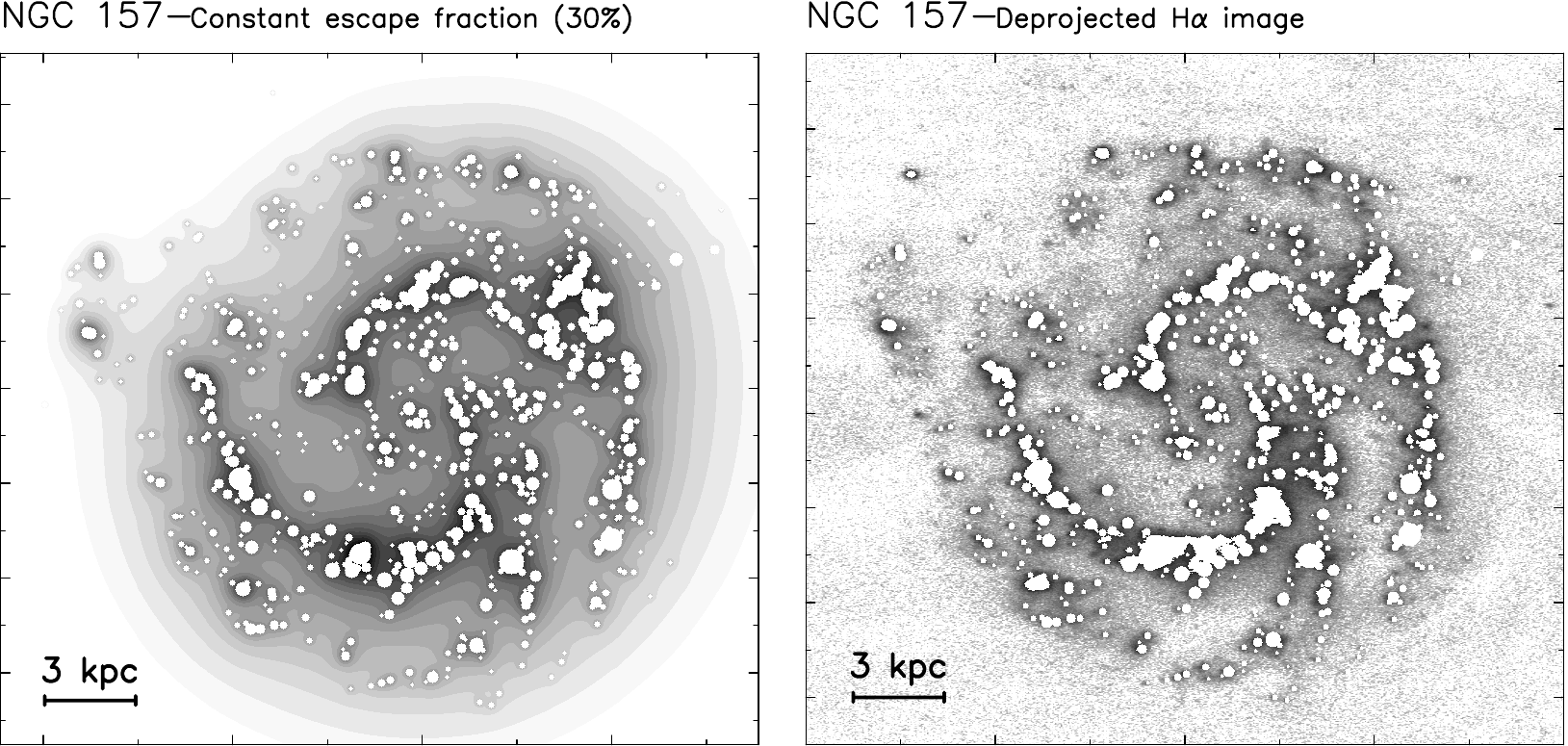}}
\caption{Comparison of a DIG model with observations. (\emph{left}) Modeled surface brightness in \ha\ of the DIG in the
    disk of NGC\,157 assuming 30\% of Lyc photons escape from each \hii\
    region,
    and a simple propagation law through a (macroscopically) uniform
    slab model for the disk. The result is a projection in the plane of
    the predicted 3D \ha\ column density. (\emph{right}) Deprojected image of 
    the galaxy with \hii\ regions masked, and a cut--off of $73\times 
    \cos i$ pc cm$^{-6}$
    applied to limit the \hii\ region contamination of the DIG. This shows
    the basic similarity between these models and the observed DIG,
    though further refinements are important (see Fig.~\ref{fig:beckman6}). From 
    \citealp{ZBR02}.
}
\label{fig:beckman5}
\end{center}
\end{figure*}

   Given a full catalog of \hii\ region positions and luminosities
   for a galaxy, one can test the hypothesis that escaping photons from
   the \hii\ regions cause the ionization of the DIG by modeling the
   transfer of these photons from their points of origin. This was done
   in considerable detail by \citet{ZBR02} for NGC\,157. This
   galaxy was selected because of the availability of a VLA \hi\ map of
   reasonable resolution, as we will explain shortly. In 
   Fig.~\ref{fig:beckman5}, we show
   a comparison between the observed surface brightness distribution in
   the DIG and one of the simplest models used. In this model 
   30\% of
   emitted Lyman continuum photons escape from each \hii\ region, and 
   propagate through
   the DIG isotropically. 

\begin{figure}[tbp]
\begin{center}
\resizebox{\linewidth}{!}{\includegraphics{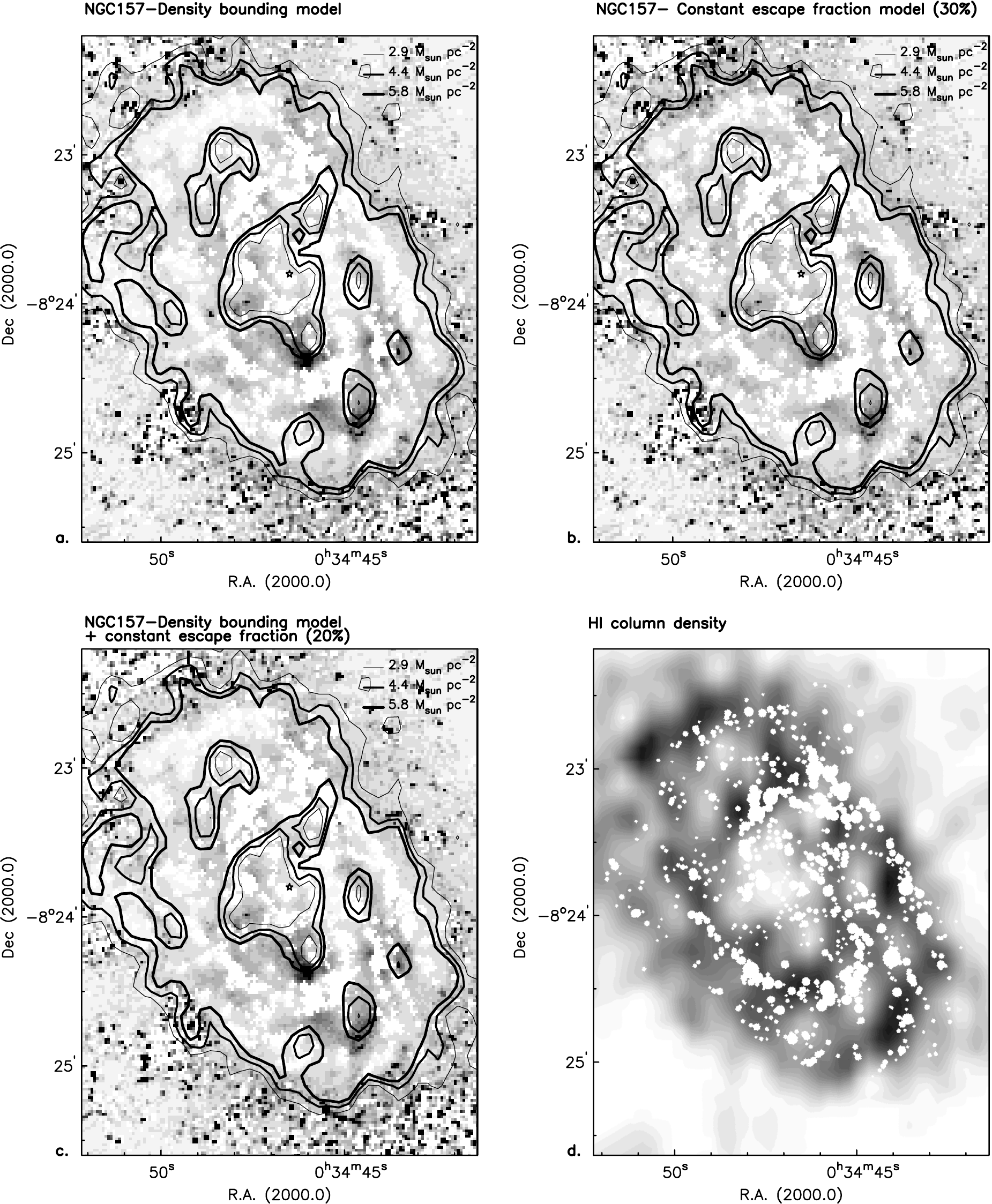}}
\caption{Ratios of observed to modeled DIG of NGC\,157, for
    three variant models of photon escape from \hii\ regions, assuming
    a slab structure to the \hi\ in the disk. (\emph{a}) Constant fraction of
    photons emitted by the O~stars escape from each \hii\ region. (\emph{b}) Only
    \hii\ regions with luminosities higher than a critical value show
    significant Lyc photon escape. (\emph{c}) A constant underlying escape fraction
    for all \hii\ regions plus an increasing increment for \hii\ regions
    above a set \ha\ luminosity. The results are similar for all three.
    (\emph{d}) Observed \hi\ column density map (with catalogued \hii\ regions
    overlaid in black). Note complete coincidence of low zones of
    observed/modeled ratio with zones of low \hi\ column density, as
    predicted if photon escape from the \hii\ regions is the main
    ionizer of the DIG. From \citealp{ZBR02}.}
\label{fig:beckman6}
\end{center}
\end{figure}

The predicted \ha\ surface brightness is derived
   by  summing the  contributions to the ionizing radiation field from
   each of the \hii\ regions.
   We can
   see that the result is remarkably similar globally to the observed
   distribution, and is itself a fair verification of the initial
   hypothesis. However a more quantitative look at the comparison shows
   that the ratio between the predicted and observed DIG surface
   brightness is not uniform on large scales, as would be expected
   since the initial model assumes a uniform slab structure for the \hi\
   involved in converting
   the Lyman continuum photons to \ha. The missing structural parameter
   can be supplied by using
   the observed \hi\ column density, as shown in
   Fig.~\ref{fig:beckman6}, where we can see
   that in zones of low \hi\ column density the ratio of observed to
   predicted DIG surface brightness is reduced. Maps of these two   
   quantities give excellent coincidence of features, and go a step
   further in showing that the principal  DIG ionization sources must be  
   the O~stars in luminous \hii\ regions. Modeling the effect of clustered 
   supernovae on the distribution of the \hi, \citet{CO02} also 
   found that the resulting clumpiness of the medium had a 
   significant effect on the escape fraction of the ionizing radiation.  
   \citet{ZBR02} carried out
   a number of different modeling tests of the basic hypothesis, varying
   the law relating the escape fraction of ionizing photons with
   the
   luminosity of an \hii\ region, and varying the mean absorption
   coefficient of the inhomogeneous neutral fraction of the DIG. However,
   they concluded that it was not possible without \hi\ data of improved
   angular resolution to go further in testing different photon escape
   laws, or to estimate what fraction of the DIG ionization
   could be due to mechanisms other than that tested.

\subsection{Line ratio studies}
\label{usingratios}

\begin{figure}[tbp]
\begin{center}
\resizebox{\linewidth}{!}{\includegraphics{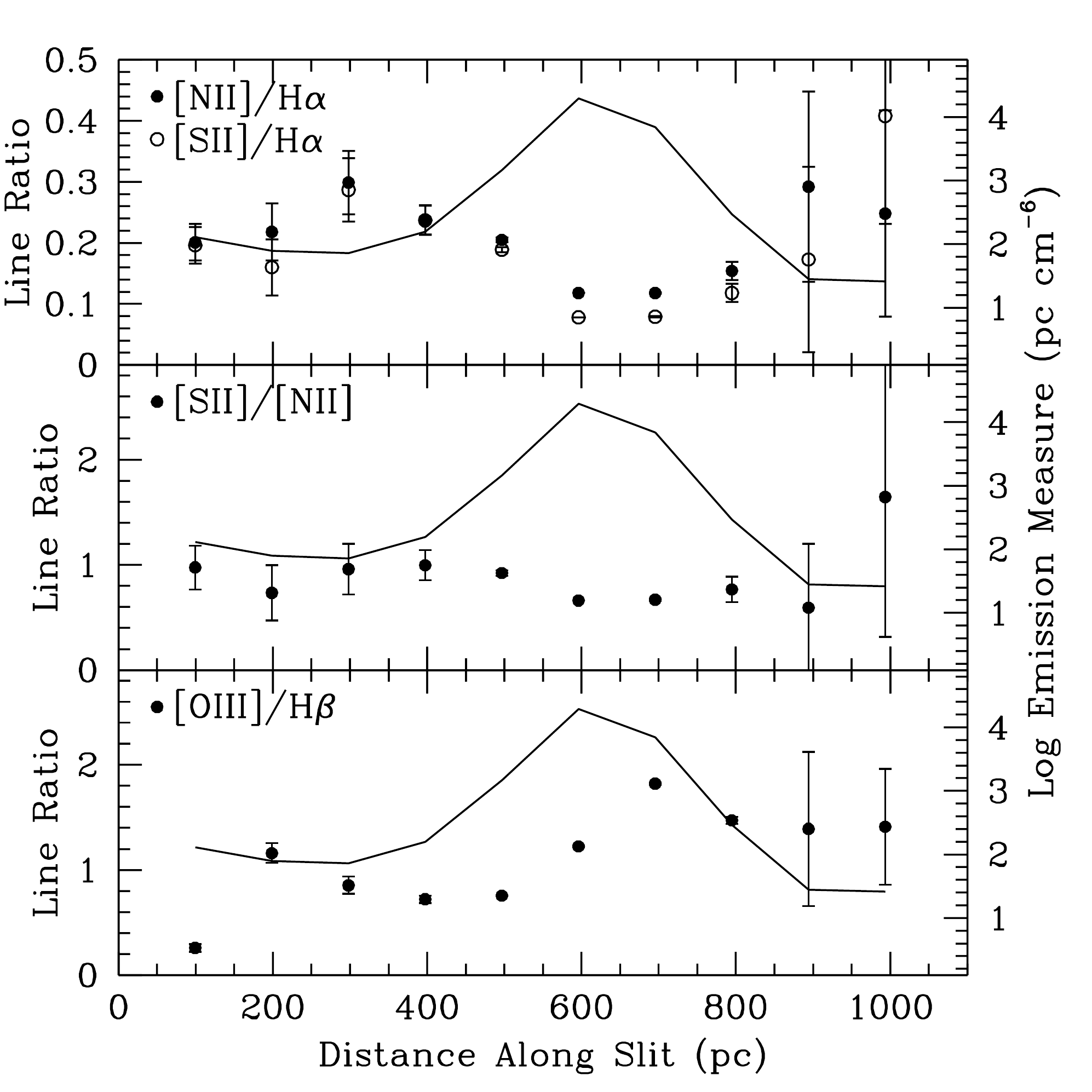}}
\caption{Selected emission line ratios from long slit spectra across
    the luminous \hii\ region NGC\,604 in M33. The solid line is the \ha\
    surface brightness showing rising to the center of the \hii\ region,  
    and falling to low values in the DIG outside it. Note the tendency of the
    line ratios to rise in the DIG.
    From \citealp{HW03}.}
\label{fig:beckman7}
\end{center}
\end{figure}

There have been relatively few quantitative spectroscopic studies of the 
DIG in face--on galaxies.  A pioneering study of \nii\ and \ha\ across the 
face of NGC\,1068 was carried out by \citet{BSC91}, who found very high 
\nii/\ha\ ratios and discussed possible causes for this high excitation.  
\citet{HW03} and \citet{VW06} have made the most careful and detailed 
examinations to date. In Fig.~\ref{fig:beckman7}, we show the 
observations of the DIG close to the luminous \hii\ region NGC\,604 in the 
nearby spiral M33 \citep{HW03}, obtained by placing a slit across the 
\hii\ region so that it sampled the DIG on either side. This example is 
representative of their study of the three local face--on galaxies, M33, 
M51, and M81. We can see that the line ratios, \nii/\ha, \hei/\ha, and 
\oiii/\hb, all tend to show higher values in the DIG. The full study 
includes measurements of these ratios as well as \sii/\ha\ and \sii/\nii\ 
in \hii\ regions and in the DIG, both in the spiral arms and in the 
interarm zones in each galaxy. The results were compared with predictions 
made with ``standard" \hii\ region ionizing fields, with varying 
ionization parameters and stellar photospheric temperatures between 
20~000~K and 50~000~K.  There is good agreement for the \hii\ regions but 
poor agreement for the DIG. They then considered more realistic radiation 
fields, which take into account the fact that the spectrum of the 
radiation escaping from \hii\ regions may be different from that from a 
pure O star (see next section).  They modified their spectra accordingly, 
varying the modeled escape fraction between 30\% and 60\%, where a lower 
escape fraction implies a harder spectrum. These results showed better 
agreement with the DIG observations than the previous set, but the 
agreement was only fair.

The authors reached the tentative conclusion that O star photoionization 
is not the sole mechanism for ionizing and/or heating the DIG, the same 
conclusion reached by others who have studied these line ratios both in
our Galaxy and others \citep[e.g.,][]{WM04, RHT99, Rand98}. This 
may indeed well be the case; however, before reaching a definitive 
conclusion, it would be useful to test a modification of the types of 
models proposed in \citet{HW03} based on the assumption that not only is 
the DIG itself inhomogeneous, but so also are the \hii\ regions. The 
effects of clumping of the interstellar gas on the escape fraction and the 
spectrum of the ionizing radiation is discussed in more detail below.

\section{Modeling the WIM/DIG: Effects of radiation transfer through a clumpy interstellar medium}
\label{sec:clumpy}

Models of a more realistic (i.e., 3-D clumpy) medium may provide a clearer 
view of the ionization and heating processes in the gas as well as insight 
into such questions as:

\begin{itemize}

\item
What are the constraints on the structure of the interstellar
medium
that allow photons to penetrate from the midplane O~stars to large
distances in the halo?

\item
How much of the \ha\
comes from the ionized surfaces of dense clouds and how much from a 
smoother low density medium occupying the space between the clouds?
Is a cloud-intercloud model sufficiently realistic?

\item
How much extra, non-ionizing heating is necessary to explain some of 
the line ratios?

\item
What is the role of interfaces between the ionized and neutral gas?  Are
photoionization codes missing crucial physics at interfaces? 

\item
How much
ionizing radiation escapes from galaxies, and what is the spectrum of
the escaping radiation?

\end{itemize}

\subsection{What exactly is an \hii\ region?}
\label{whatshii}
    
Before describing various modeling efforts for the DIG, we will try to 
envision the structure of an \hii\ region.  In the classical picture of a 
Str{\"o}mgren sphere, a spherical volume of interstellar gas of uniform 
density is ionized by a central source. For plausible values of the 
ionization parameter there is a sharp transition from \hii\ to \hi\ at the 
Str{\"o}mgren radius \citep{str39, o89}. However, in order to better 
explain line ratios in real \hii\ regions, the ``filling factor" approach 
was devised in which the ionized gas is confined to small, fully ionized 
clouds surrounded by vacuum, and with the ionized gas occuping only some 
fraction of the \hii\ region volume \citep{str48, of59, o89}.  Both of 
these pictures deviate from real \hii\ regions, where neutral 
condensations can exist within the \hii\ region.  In addition there may be 
a convoluted interface between ionized and neutral gas on the faces of 
clouds exposed to ionizing radiation. These interfaces and the shadowed 
regions behind the clumps will have a very different ionization and 
temperature structure and hence also a different emission line spectrum 
compared to fully ionized clouds or the gas inside a uniform Str{\"o}mgren 
sphere \citep[e.g.,][]{Williams92}.  Therefore, a more realistic picture 
of an \hii\ region comprises high density clumps embedded in a lower 
density interclump gas.

\begin{figure}[tbp]
\begin{center}
\resizebox{\linewidth}{!}{\includegraphics{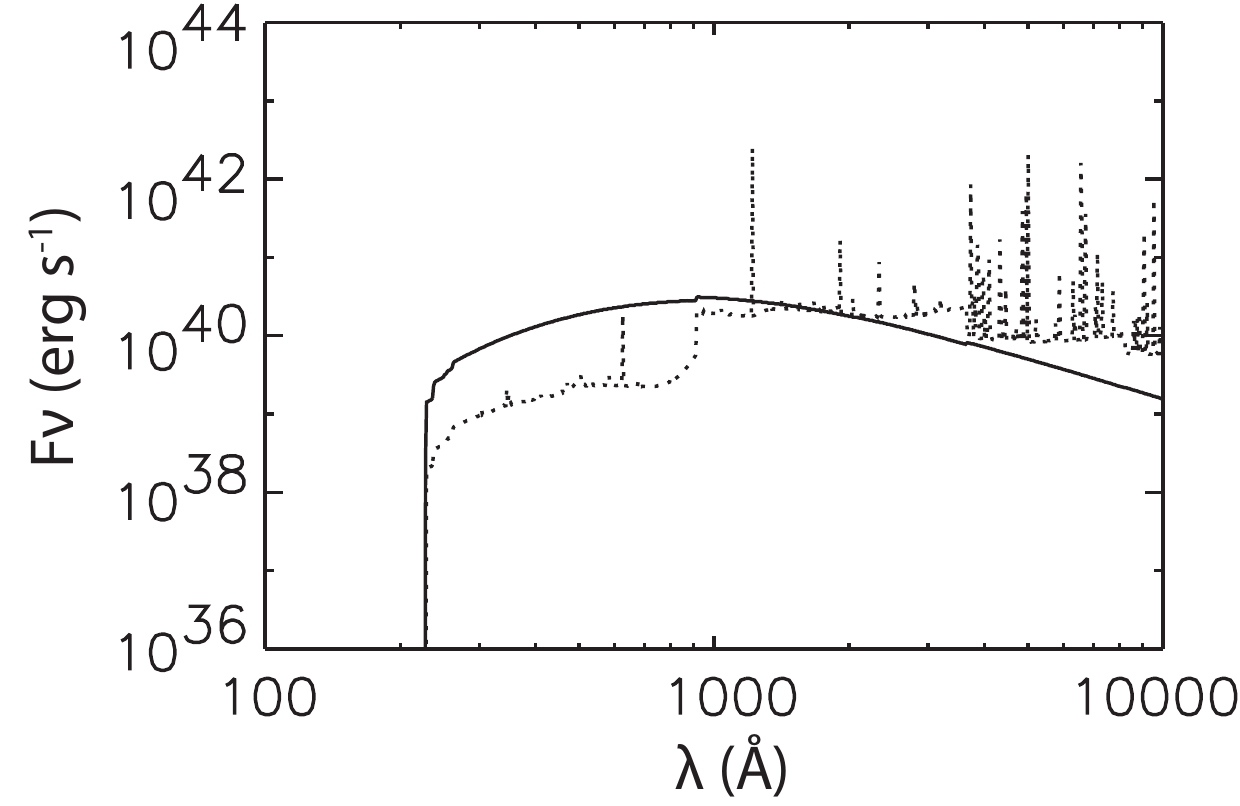}}
\caption{Components of unattenuated and hardened ionizing spectra
    emerging from a clumpy model \hii\ region, with geometrical filling  
    factor of denser clumps 10$^{-2}$, clump density 100~cm$^{-3}$, and  
    mean
    clump radius 1~pc. Solid line: input spectrum emerging diluted but
    unchanged in relative intensities through interclump medium. Dashed
    line: hardened spectrum after transiting internal clumps. In this
    model the DIG would be ionized by a combination of both spectra,
    weighted according to the details of the clump model. For details
    of these models see \citealp{giam04}.}
\label{fig:beckman8}
\end{center}
\end{figure}

Depending on the density of the clumps they may be fully ionized or be 
dense enough so that they have an ionized skin and a shielded, neutral 
interior \citep[e.g.,][]{giam04}.  This more complex picture can 
qualitatively explain the observed features of \hii\ regions including 
``blister" regions such as the Orion Nebula \cite[e.g.,][]{fer01}. 
Consequences of this clumpy structure for emission line ratios from \hii\ 
regions were quantified in \citet{giam04} and in \citet{GBC05}. Based on 
the models explained in these two articles, Fig.~\ref{fig:beckman8} 
shows an example of two spectra of ionizing photons emerging from an \hii\ 
region, one hardened by partial ionization of \hi\ while propagating 
through the denser clumps within the region, and the other essentially 
unaffected as it has passed through the interclump gas. Applications to 
the DIG will entail the combination of these two types of 
spectra with suitable weightings. When this has been done, we will be in a 
position to make clearer statements about what fraction of the ionization 
and heating of the DIG can be due to photons escaping from \hii\ regions, 
the presumed primary source of the ionization.

\subsection{Modeling the ionization structure of the DIG}
\label{modelingstructure}

A complex 3D picture is almost certainly needed to explain the DIG. As we 
will see below, for example, a 3D medium can help to answer questions 
regarding the penetration of ionizing photons to large distances. However, 
among the most crucial parts of any 3D photoionization model are the 
density distributions of its gaseous components. If, for example, we adopt 
a model consisting of dense \hi\ clouds within a uniform lower density 
intercloud \hi\ medium, then we will need to know what is the density of 
the intercloud medium, how much does it contribute to the observed 
H$\alpha$, and what is the covering factor of the clouds?  In models with 
more complex density distributions, we would need more accurate knowledge 
about the actual spatial and density distributions of the ionized 
as well as the neutral gas, which are only beginning to be explored 
\citep[e.g.,][]{Hill08, KRK08}.  Questions pertaining to the density 
structure of the interstellar medium are the major unknowns in 
photoionization models.

Until recently, multi-dimensional photoionization models of the DIG did 
not consider the faint forbidden line emission, and focussed mostly on 
H$\alpha$ and the 3D ionization structure of hydrogen via ``Str{\"o}mgren 
volume'' techniques \citep{franco90, MC93, ds94, dsf00} or Monte Carlo 
simulations \citep{wl, c02}.  These models demonstrated that extended DIG 
layers could be produced by the ionizing radiation of O stars in a clumpy 
medium, where the 3D density structure of the medium was simulated using a 
variety of models including ``standard clouds'' \citep{MC93}, superbubbles 
formed by the action of supernovae and stellar winds \citep{dsf00}, and 
two-phase and fractal densities \citep{wl, c02}.  A 3D density structure 
was found to be necessary to allow ionizing photons to penetrate to large 
distances from the O stars, which are confined to discrete 
locations near the galactic midplane.  In a smooth medium with a typical 
interstellar density of 1~cm$^{-3}$, an O star will form a Str\"omgren 
Sphere of radius about 60 pc, and within the neutral medium an ionizing 
photon will penetrate only 0.1~pc; so, 3D structures must provide the low 
density paths that allow the photons to traverse kiloparsec size-scales 
and ionize the gas far from the O~stars and at large heights above the 
midplane.
 
\subsubsection{Escape of ionizing radiation through superbubbles}

The work of \citet{dsf00} investigated whether superbubbles created by the 
dynamical action of supernovae and stellar winds could provide the low 
density paths.  They found that the dense swept up shell of material would 
trap ionizing photons within the bubble and thus inhibit the escape of 
ionizing radiation to the halo, unless, of course, the size of the bubble 
were so large that it reached into the halo.  Their models employed a 
smooth, continuous structure for the swept up shell.  The fragmentation of 
shells and the punching of holes by supernovae and stellar winds as they 
expand into a clumpy medium may offset the trapping of radiation in smooth 
models and allow a larger fraction of ionizing photons to escape.  Recent 
analysis by \citet{ter2003} of the W4 chimney suggests that a significant 
fraction of ionizing radiation is able to escape from a fragmented 
shell-like structure near the midplane, perhaps to be absorbed on the 
distant wall of a much larger cavity that reaches into the lower halo 
\citep{Reynolds+01b}.  Photoionization models of bubbles created by 
stellar winds in a turbulent star forming cloud also suggest that large 
escape fractions (in excess of 20\%) are possible \citep{dale05}. Future 
work should study in more detail the ionization structure of 3D 
hydrodynamic simulations of supernova created bubbles and superbubbles to 
determine escape fractions and the spectrum of the escaping ionizing 
radiation.  A comparison of such models with observations of \hii\ regions 
and the surrounding DIG would allow a critical test of the predicted 
interstellar structure.

\subsubsection{Two- and three-dimensional ionization structure of the DIG}

The \citet{MC93} model of the DIG adopted a two component, vertically 
stratified density structure $n({\rm H}) = 0.1\exp(-|z|/0.3) + 
0.025\exp(-|z|/0.9)$, where the number density is per cm$^3$ and the 
z-distance is in kiloparsecs.  This represents the concentrated neutral 
layer and the extended ionized layer.  They also included an approximation 
for absorption by dense opaque clouds using a model that reproduces the 
statistics of clouds in the interstellar medium.  Although their smooth 
density is lower than the average density inferred for $\mathrm{H}^{0}$ 
and $\mathrm{H}^{+}$, when implemented in their ``standard cloud'' model 
including the known ionizing sources in the solar neighbourhood, it 
reproduced the average observed emission measures (EM) and dispersion 
measures (DM).

An obvious criticism of the Miller \& Cox model is that the mean density 
they used is smaller than that inferred for the H~I in the Galaxy 
\citep[e.g.,][]{dl90}.  However, they did use the known distribution and 
ionizing luminosity of O stars in the Solar neighborhood and showed that 
the above density distribution would allow for the gas to be ionized to 
large $|z|$.  In reality the gas is clumped on a wide range of size 
scales, so the Miller \& Cox density provides an estimate of the maximum 
amount of smooth component allowed, if ionizing photons are to penetrate 
to large $|z|$.  Therefore, the Miller \& Cox density profile may be a 
good starting point for the smooth intercloud medium in 3D models that 
incorporate both smooth and clumpy components.

\begin{figure}[!tb]
\resizebox{\linewidth}{!}{\includegraphics{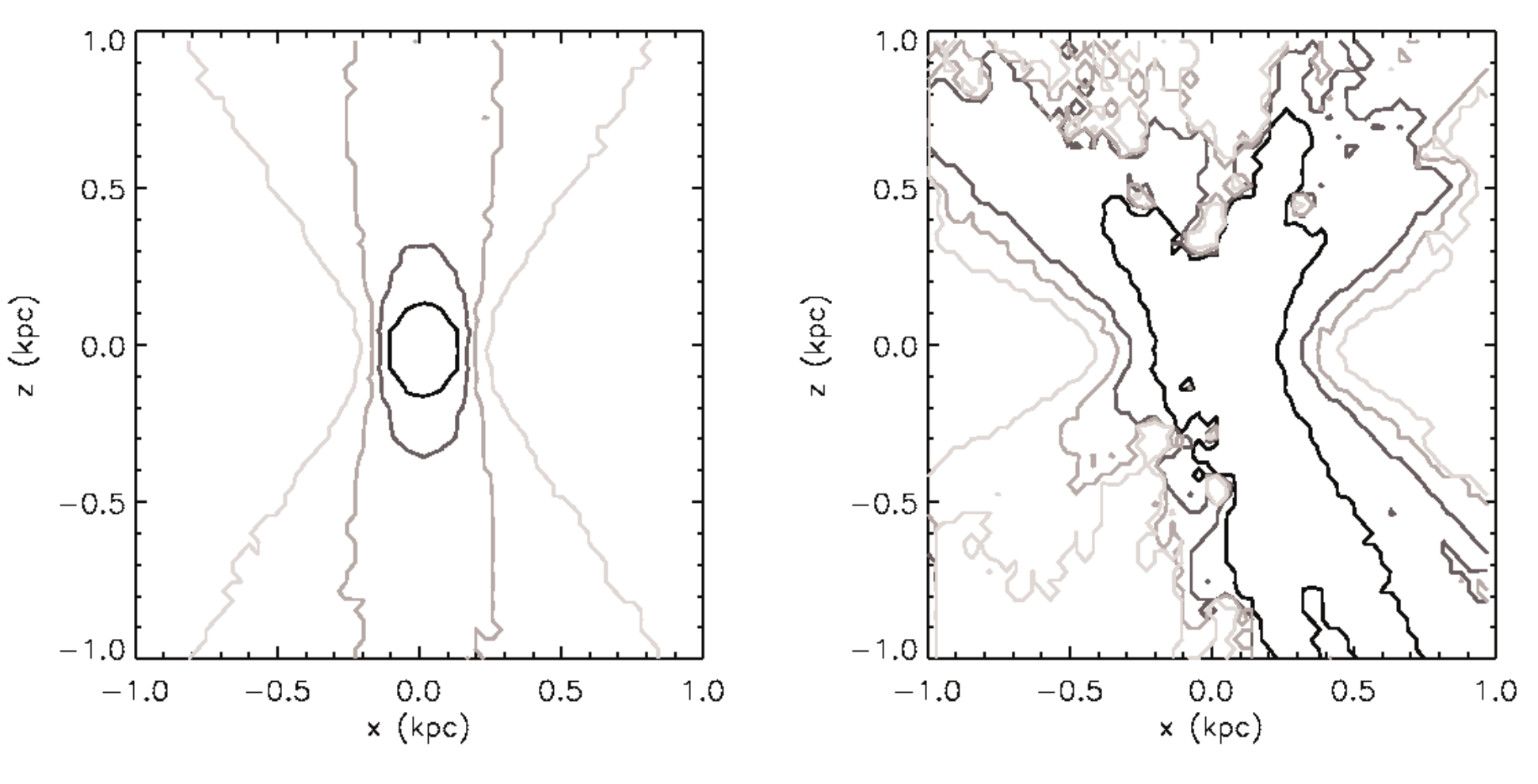}}
\caption{Slices through the ionization structure for point sources in 
(\emph{left}) smooth and (\emph{right}) hierarchically clumped density distributions.  
The mean vertical density structure for each simulation is that of a 
Dickey-Lockman disk.  From the inner to outer contours the source 
luminosities (photons per second) are $10^{49}$, $3\times 10^{49}$, 
$5\times 10^{49}$ and $10^{50}$.  Comparing the contours in the smooth and 
clumpy models, it is clear that the 3D density structure provides 
low density paths allowing photons to reach much larger distances from 
the ionizing source.}
\label{fig:wood1}
\end{figure}

To illustrate the effect of a clumpy, 3D density distribution, 
Fig.~\ref{fig:wood1} shows slices through the ionized volumes created in 
a completely smooth medium and in a medium with both clumpy and smooth 
components.  In each case to simulate an O~star, the ionizing source is a 
40~000~K blackbody point source at the center of a grid which is 2000~pc 
on a side.  The hydrogen ionizing luminosity Q is varied in the range 
$10^{49}\,{\rm s}^{-1} < Q < 10^{50}\,{\rm s}^{-1}$, thus giving a range 
of sizes for the ionized volumes, corresponding to that for a single 
O~star up to that for a large O~star association.  For the smooth 
medium the density structure is that of a Dickey-Lockman disk 
\citep{dl90}.  For the clumpy medium the average $|z|$-height dependent 
density is also that of a Dickey-Lockman disk, except that the gas has 
been turned into 3D clumps using the hierarchical clumping algorithm of 
\citet{e97}. For this simulation one third of the mass remains in a smooth 
component and the remainder is converted into hierarchical clumps.  The 
photoionization simulation is performed using the 3D Monte Carlo code of 
\citet{wme04}.

The contours in Fig.~\ref{fig:wood1} show slices through the ionized 
regions for different source luminosities.  It is immediately seen that 
for a given source strength, the 3D density structure with clumps allows 
ionizing photons to reach much larger $|z|$ distances than the completely 
smooth distribution.  In the smooth model the source with 
$Q=10^{49}$~s$^{-1}$ (typical of a single O star) creates a very small 
ionized volume, while the same source ionizes a much larger volume in the 
3D simulation.  Recall that both the smooth and clumpy density grids have 
the same mean $n(z)$ given by the Dickey-Lockman density distribution.  
Therefore, it appears that 3D density structure readily allows for 
ionizing photons to penetrate to large $|z|$, thus solving a problem 
common to models that adopt a smooth density structure of the medium.

\subsection{Modeling the emission line spectrum of diffuse ionized gas}
\label{modelingspectrum}

The physical conditions in the DIG are revealed through its emission 
line spectrum, which probes its ionization state and temperature 
structure (\S \ref{sec:diving}).  The main observational 
characteristics that successful models of the DIG spectrum must 
address are:

\begin{itemize}

\item 
The temperature, probed via the dominant ion ratios [N {\sc 
ii}]/H$\alpha$ and [O~{\sc ii}]/H$\alpha$, appears to rise with increasing 
$|z|$-height above the plane.  Most of the nitrogen and oxygen are N$^+$ 
and O$^+$, and [N {\sc ii}] and [O {\sc ii}], excited by electron 
collisions, are very sensitive to temperature.

\item 
Compared to traditional \hii\ regions the DIG appears to be 
ionized by a softer radiation field.  This is revealed in the elevated 
[S {\sc ii}]/[N {\sc ii}], a measure of S$^+$/S, and the apparent 
underionization of He implied by low He~{\sc i}/\ha\ 
recombination line intensity ratios.

\item 
WHAM observations show that [O {\sc i}]/H$\alpha$ is weak, at least 
near the Galactic midplane, while in NGC\,891 [O {\sc i}]/H$\alpha$ is 
observed to increase with increasing distance above the plane. Models 
predict that interfaces between the ionized and neutral regions should be 
revealed by increased [O {\sc i}]/H$\alpha$ and a rapid rise of [N {\sc 
ii}]/H$\alpha$ and [S {\sc ii}]/H$\alpha$ near the transition region, 
where the gas temperature is rising rapidly.

\item 
In some galaxies [O {\sc iii}]/H$\alpha$ is seen to remain high or 
even increase at large $|z|$-heights, whereas photoionization models 
predict [O {\sc iii}]/H$\alpha$ to decrease with height above the plane.

\end{itemize}

\subsubsection{One-dimensional models}

While the 2D and 3D techniques described above were used to explain the 
{\it formation and structure} of the DIG via O-star photoionization, 
detailed 1D photoionization codes were used to explain the DIG's {\it 
emission line spectrum} \citep[e.g.,][]{m86a, dm94, m00, SHR2000, cr01, 
HW03}.  In general, the 1D simulations parameterized the problem with an 
ionization parameter, calculated the ionization structure of a constant 
density spherically symmetric volume, and formed line ratios by taking 
ratios of the total emission line intensities from the entire volume.  By 
combining simulations with different ionization parameters and different 
supplemental heating rates, these volume averaged models could explain 
many of the observed line ratios in the DIG.  However, 1D models do not 
allow for a self-consistent explanation of the observed {\it trends} of 
line ratios.  In particular, [N~{\sc ii}]/H$\alpha$ and [S~{\sc 
ii}]/H$\alpha$ are observed to be anticorrelated with the H$\alpha$ 
intensity and increase with increasing distance from the midplane 
\citep[e.g.,][]{HRT99, Rand98}.

Explaining these observational trends requires models that do not use 
volume averages, but form line ratios for lines of sight that pierce 
through the ionized volume at increasing impact parameters away from the 
ionizing source.  The plane parallel slab models presented by 
\citet{bfq97} showed how the depth dependence 
of the ionization and temperature structure (see their Fig. 9) could 
explain the trends for [N~{\sc ii}]/H$\alpha$ line ratios with distance 
from the source.  More recently, \citet{ED05} presented a grid of 
plane 
parallel models that could explain the observed increase of [N~{\sc 
ii}]/H$\alpha$ and [S~{\sc ii}]/H$\alpha$ above the midplane.

A persistent problem with all these photoionization models is that they 
cannot explain the rise of [O~{\sc iii}]/H$\alpha$ with height above the 
plane unless some other ionization and/or heating source is invoked, such 
as shocks, photoelectric heating, or turbulent mixing layers 
\citep{cr01, RHT99, SSB93}.  Also, unless the medium is fully ionized 
beyond the edge of the simulation (i.e., density bounded) the models 
predict uncomfortably large [O~{\sc i}]/H$\alpha$ ratios at the edge of 
the ionized volume.  Clumping in \hii\ regions may explain the 
apparent absence of these edge effects as discussed below.

\subsubsection{Two- and three-dimensional models}

Recent advances in the development of Monte Carlo photoionization codes 
\citep{olr, erc, wme04} now enable calculations of the ionization and 
temperature structure and emission line strengths for multiple ionizing 
sources within 3D geometries.  \citet{WM04} presented simulations for 
photoionization and line intensity maps of a stratified interstellar 
medium.  Their models showed that due to hardening of the radiation field 
in the spectral region between the \hi\ and \hei\ ionization edges, the 
temperature naturally increases away from the ionizing source.  They also 
found that [N~{\sc ii}]/H$\alpha$ and [S~{\sc ii}]/H$\alpha$ rise with 
$|z|$ in their vertically stratified density structure and that [S {\sc 
ii}]/[N {\sc ii}] remains fairly constant with $|z|$ as observed.  While 
their models could explain most of the line ratios, additional 
non-ionizing heating was still required to produce the largest line ratios 
observed.

\subsubsection{Interfaces and three-dimentional \hii\ regions}

\begin{figure}[tbp]
\resizebox{\linewidth}{!}{\includegraphics{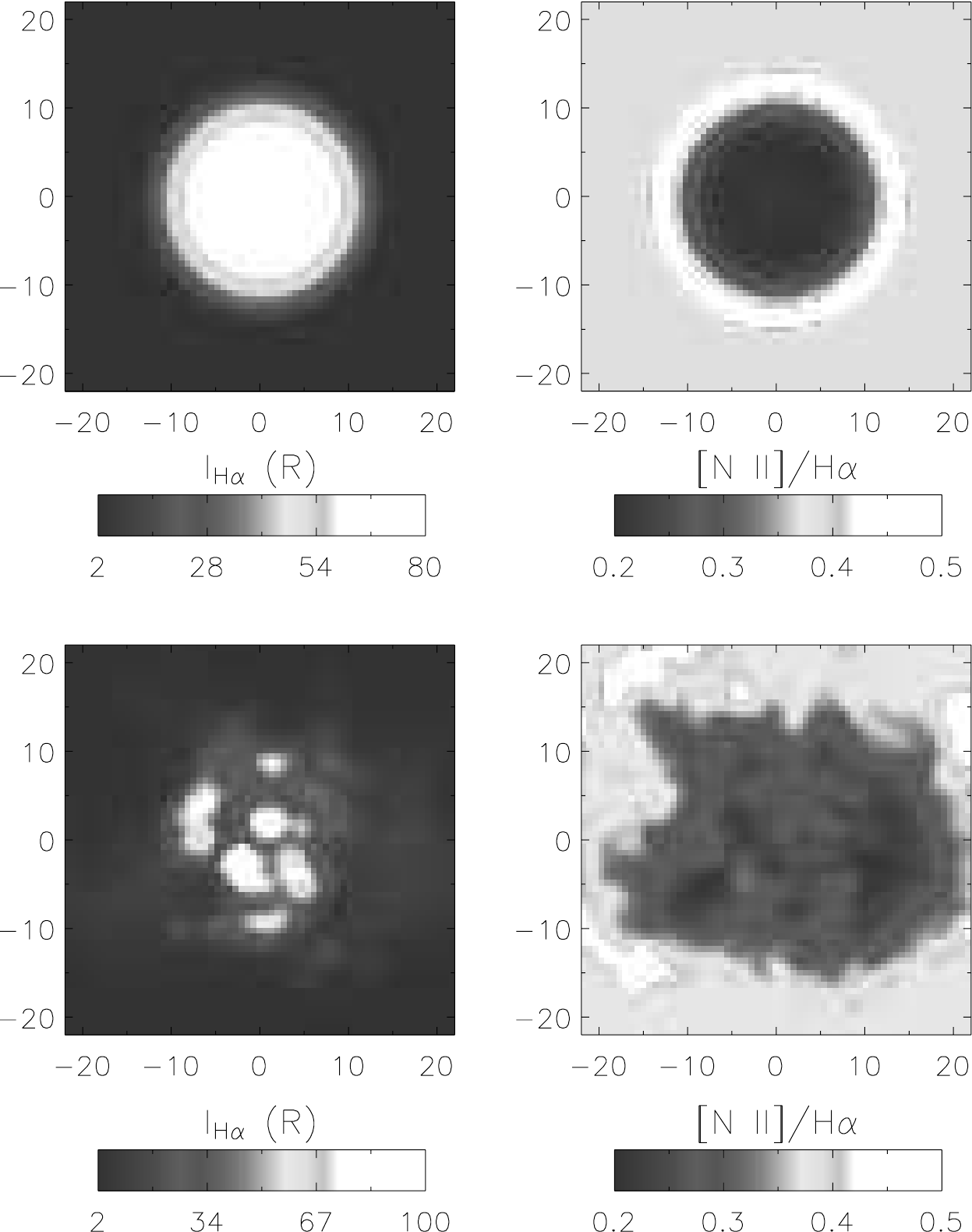}}
\caption{H$\alpha$ and [N {\sc ii}]/H$\alpha$ maps for (\emph{upper panels}) uniform density 
and (\emph{lower panels}) hierarchically clumped \hii\ 
region models.  In each simulation a uniform background has been 
added to the \ha\ and \nii\ maps to simulate the \hii\ 
region being surrounded by low density DIG.  Notice the very 
irregular boundary in the clumpy model and that the rise of 
\nii/\ha\ with distance from the central source is much 
steeper in the smooth model.}
\label{fig:wood2}
\end{figure}

In standard photoionization models of the interface between ionized and 
neutral gas, the medium is turning neutral and the temperature is rising 
rapidly due to the hardening of the radiation field \citep[e.g.,][]{WM04}.  
For optical observers, such an interface should be characterized 
by enhanced emission of \oi, \nii, and \sii\ 
relative to \ha.  As mentioned above these effects have in general 
not been observed in the DIG.  However, recent results from the WHAM 
survey of low emission measure \hii\ regions do appear to show an 
increase of \nii/\ha\ and \sii/\ha\ away from 
the ionizing sources \citep{WHR05}.  Fig.~\ref{fig:wood2} shows that 
models of hierarchically clumped \hii\ regions give a shallower rise 
of \nii/\ha\ away from the ionizing source than that 
predicted by uniform density models.  This is because the edge of the 
\hii\ region is very irregular and sightlines probe different columns 
of ionized gas and different temperatures compared to the regular, rapid 
rise in temperature and neutral fraction seen at the edge of a uniform 
density \hii\ region.  This shallower rise is in better agreement 
with the observations. Further work extending 3D models from individual 
\hii\ regions to study photoionization in global models of the 
interstellar medium is required.  This will test whether interfaces can 
indeed be masked by 3D effects and the edges of \hii\ regions may be 
lost in diffuse foreground and background (non-interface) emission, 
although the observed weakness of the \oi\ emission still presents 
a problem.  Most photoionization codes neglect the effects of shocks and 
the expansion of ionization fronts at the edges of \hii\ regions.  
Perhaps there is some crucial missing physics from current photoionization 
codes that is important for modeling interface emission.  There are 
efforts underway to include such effects in the CLOUDY photoionization 
code (Elwert 2005, private communication).

\subsubsection{Leaky \hii\ regions and the He$^+$/H$^+$ problem}

Leaky \hii\ regions appear to be an important source of ionizing photons 
for the DIG (see the discussion in \S \ref{sec:faceon}).  The ionizing 
photons escape from them either directly through empty holes or via escape 
through density bounded ionized gas.  In the first case the escaping 
spectrum will be that of the ionizing star, while in the second case the 
spectrum will be modified due to partial absorption of the radiation by 
neutral H and He. Compared to the intrinsic stellar spectrum the escaping 
spectrum is generally harder in the H-ionizing continuum and has its 
He-ionizing photons supressed \citep[e.g.,][]{HW03, WM04}.

A combination of direct escape and transmission through density bounded 
ionized gas is likely to occur in a 3D \hii\ region. In fact, line 
ratios in the DIG surrounding leaky \hii\ regions may allow us to 
determine whether the escaping photons are dominated by those escaping 
through true holes in the \hii\ region or by those penetrating 
density-bounded \hii\ gas.  Specifically, the supression of 
He-ionizing photons penetrating density bounded \hii\ regions may 
explain the underionization of He in the DIG. However, more observations 
of He~{\sc i} and other lines as a function of distance from \hii\ 
regions in our Galaxy and others are 
required to test this.  It cannot yet be ruled out that most of the 
DIG ionization is produced by later O stars (which have a soft 
spectrum consistent with the He~{\sc i} observations), because the 
earliest (hard-spectrum) O stars may preferentially have their ionizing 
radiation absorbed within their parent molecular clouds.  However, 
a study of face-on galaxies suggests that these cooler O~stars do 
not provide sufficient ionizing power \citep{FWG96}.  Clearly more 
observational and theoretical investigations are required into the role of 
leaky \hii\ regions and the spectrum of their escaping ionizing 
radiation.

\section{Ionizing radiation from hot gas-cool gas interfaces}
\label{sec:interface}

Early-type stars produce a prodigious amount of ionizing radiation and are 
capable of ionizing gas well above the midplane of the Galaxy. But does 
this stellar radiation explain all of the observed $\mathrm{H}^{+}$?  For 
example, do some of the O~star photons completely escape the Galaxy to 
account for the ionized gas observed in the high velocity \hi\ clouds 
(HVCs) and the Magellanic Stream, located up to 50~000 pc from the Galaxy 
\citep{WW96, TRH98, p2003}? If so, then the \ha\ surface brightness of 
these clouds provides a direct measurement of the flux of Lyman continuum 
radiation that completely escapes the Galaxy \citep[e.g.,][]{TRH98}.  
However, because these cooler clouds appear to be immersed in a much 
hotter plasma \citep[e.g.,][]{SSW03}, we must at least consider the fact 
that hot gas-cool gas interfaces are also a source of ionizing radiation. 
Such radiation may even play an important role in the ionization of a very 
local \hi\ cloud in the vicinity of the sun.

Observations of O$^{+5}$ and detections of other ions in high states of 
ionization \citep{Savage_etal_2000, Sembach_etal_2003, 
Sembach+Savage_1992} show that interfaces between hot ($\sim 10^6$~K) gas 
and cooler ($\alt 10^4$~K) gas are widespread throughout the interstellar 
medium and Galactic halo.  The resulting intermediate temperature ($\sim 
10^5$~K) gas at these interfaces produces extreme ultraviolet ionizing 
radiation. Thus, even though O~star photons can leak through a clumpy 
interstellar medium and/or through superbubble chimneys, interfaces have 
the advantage that they exist wherever the hot gas and cooler clouds do, 
including places where ionizing radiation from O~stars does not reach. 
Although significantly weaker than the flux from hot stars, interface 
radiation may be more widely distributed, and because the emission is 
generated in a thin layer adjacent to the absorbing cloud, interface 
radiation is efficiently used for ionizing that cloud as well as being a 
source of ionization for other more distant clouds.

\subsection{Types of interfaces}

\begin{figure}[tbp]
\begin{center}
\resizebox{\linewidth}{!}{\includegraphics{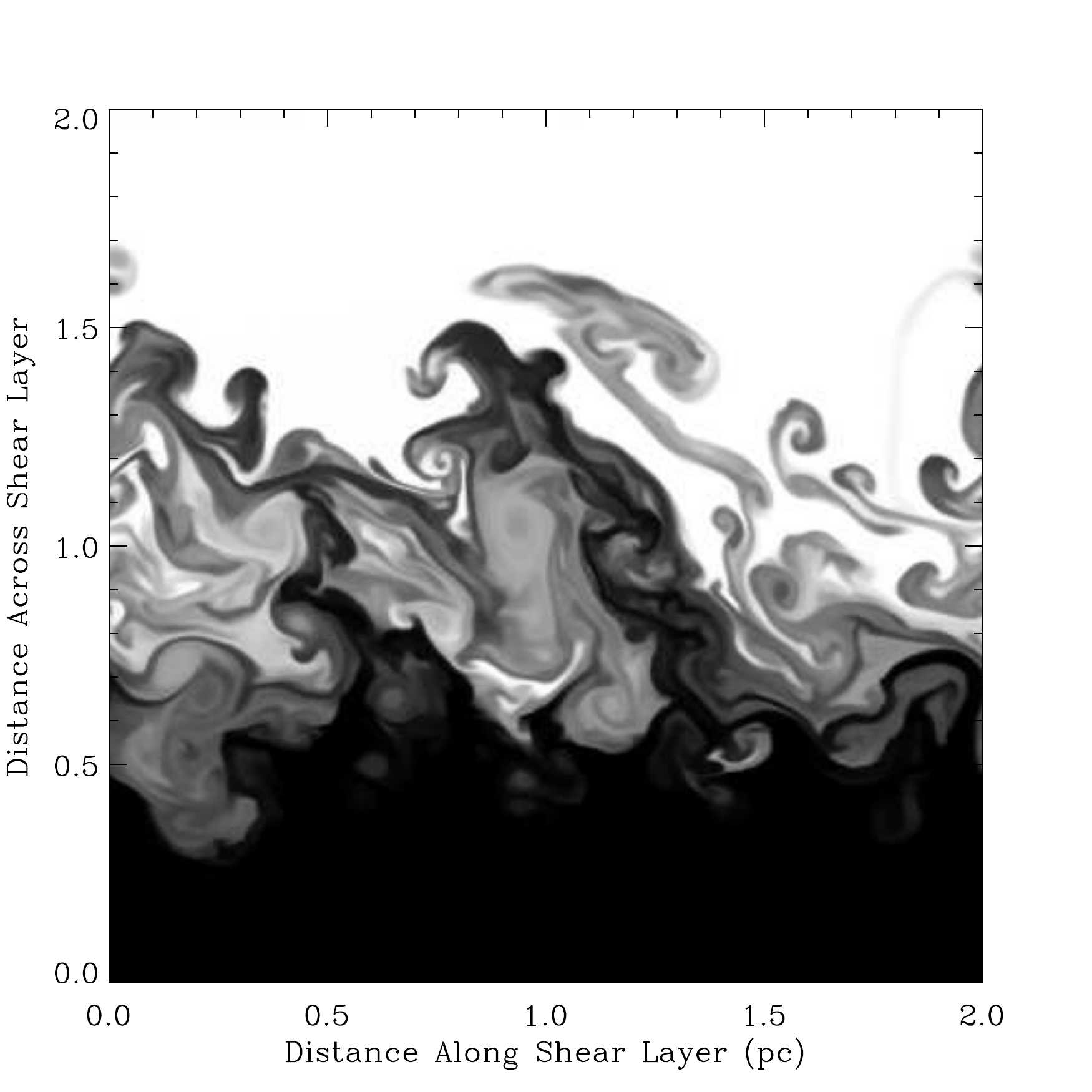}}
\caption{Temperature in a shear layer between hot and warm gas.  The hot
gas (\emph{top}, $T = 10^6$~K) flows to the left at 10 km s$^{-1}$ while the warm 
gas (\emph{bottom}, $T = 8000$~K) flows to the right at 10 km s$^{-1}$. Cooling 
is not included in this 2-D calculation.}
\label{fig:tml}
\end{center}
\end{figure}

A variety of different types of interfaces may exist, depending on the 
physical processes operating and dynamical state of the boundary region 
between the hot and cooler gas.  These include: evaporative boundaries 
\citep[e.g.,][]{CM77}, cooling/condensation fronts \citep{SB91}, and 
turbulent mixing layers \citep{SSB93, Begel+Fabian_1990}. In evaporative 
boundaries, thermal 
conduction heats the cool cloud and produces an outflow.  This requires 
that the magnetic field topology is such that the warm gas is not shielded 
too thoroughly from the hot gas. In cooling/condensation fronts, slow 
accretion of hot gas onto the cool gas occurs as the hot gas cools 
radiatively.  A turbulent mixing layer (TML) can develop in regions where 
there is shear flow at the hot/cool boundary that leads to hydrodynamical 
instabilities and mixing of the hot and cool gas.  The mixed gas in a TML 
cools rapidly due to its temperature and non-equilibrium ionization state.
Although all these
types of interfaces share some characteristics, the ionization state of
the gas can be radically different for different types of interfaces.  
For example, relative to collisional ionization equilibrium the gas can be
highly underionized in evaporative outflows, overionized in
cooling/condensation fronts, or a combination of both overionized and
underionized as in a TML, wherein the formerly cool gas is underionized
and the formerly hot gas is overionized.
In Fig.~\ref{fig:tml} we show the results of preliminary numerical 
hydrodynamic simulations of a TML. Other possibilities exist for hot 
gas/cool gas interfaces involving various combinations of cooling, 
conduction and mixing, but those have yet to be explored.

\begin{figure}[tbp]
\resizebox{\linewidth}{!}{\includegraphics{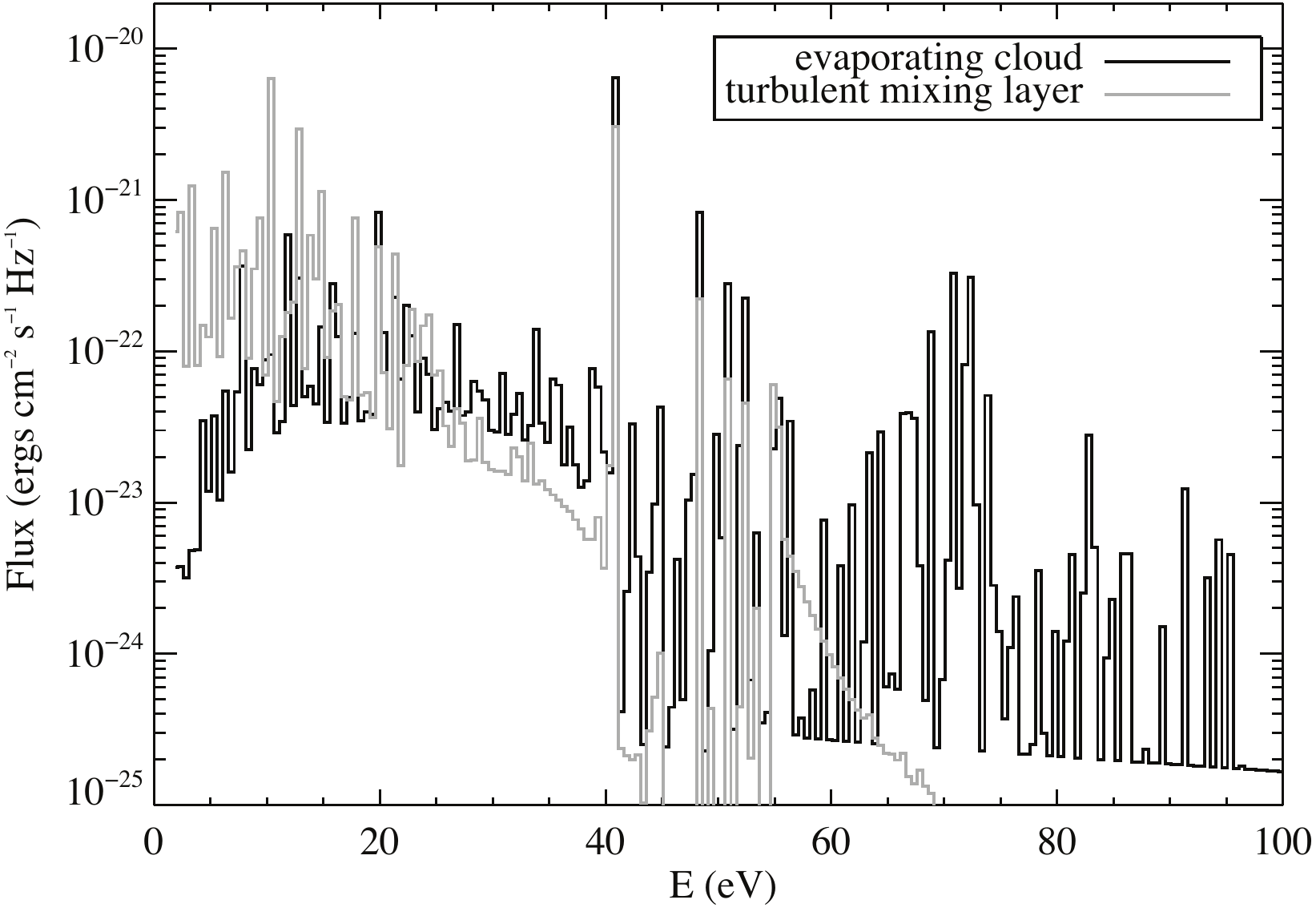}}
\caption{Comparison of ionizing flux generated in evaporating cloud with
that generated in a turbulent mixing layer.}
\label{fig:spect}
\end{figure}

The flux and spectrum of ionizing radiation emitted, is determined by the 
ionization-temperature-density profile in the interface.  In 
general, underionized gas radiates more strongly than overionized gas, 
because ions that are being ionized up will generally be excited several 
times before being ionized. In Fig.~\ref{fig:spect} we show a comparison 
of the EUV/soft X-ray spectrum generated in an example evaporating cloud 
boundary and TML. In this calculation the hydrogen ionizing photon 
production rate between 14 eV and 24 eV is approximately $2 \times 10^4$ 
photons cm$^{-2}$ s$^{-1}$.  This is only about 10\% the ionizing flux 
that appears to be incident on high velocity clouds in the Galactic halo 
\citep[e.g.,][]{TRH98, Tea02}, for example; however, the uncertainty in 
the properties and morphology of actual cloud interfaces (e.g., the number 
of interfaces a line of sight through a cloud intersects; see Fig. 
\ref{fig:tml}) leaves open the possibility that the H$\alpha$ produced by 
interface radiation could be significant in regions where stellar ionizing 
photons do not reach.

\subsection{A test case} 

The Local Interstellar Cloud (LIC) that surrounds the Solar System appears 
to be an excellent candidate for exploring interstellar interfaces.  It is 
inside the Local Bubble \citep{CR87}, and thus probably 
surrounded by hot gas. There is no direct Lyman continuum radiations from 
O~stars and the nearest B star ($\epsilon$ CMa) is more than 100 pc away. 
The ionization of the LIC is very well characterized (it is best observed 
interstellar cloud in the Universe), and its ionization state is somewhat 
unexpected, that is, quite different from the WIM.  In the LIC, the 
hydrogen is only moderately ionized at $\sim$20-40\%, and He is more 
ionized than H.

Models that include ionizing radiation 
from an evaporative interface \citep{SF02, Frisch+Slavin_2003}, are generally successful in matching the 
myriad of data available and indicate that a diffuse EUV source above the 
weak ionizing flux provided by nearby stars \emph{is} necessary to explain 
the ionization.  Problems remain, however, in explaining the relatively 
low column densities of O$^{+5}$ and C$^{+3}$ as well as the high column of 
Si$^{+2}$ that have been observed \citep{GJ01, Oegerle_etal_2005}. These 
discrepancies would seem to point to a different type of interface 
surrounding the LIC. The existence of the LIC does raise the question of 
how common such partially ionized clouds are and how much they contribute 
to the diffuse interstellar $\mathrm{H}^{+}$ \citep[e.g.,][]{Reynolds04}.  
While the density and temperature of the LIC are very close to what is 
found in the WIM, the low values of [O~I]/H$\alpha$ observed by WHAM imply 
that $\mathrm{H}^{+}$/H for most of the diffuse ionized gas in the solar 
neighborhood is much closer to unity than the lower value found for the 
LIC (although He$^+$/He may be similar).

In summary, while many aspects of the physics of interfaces are yet to be 
explored, the fluxes produced by such interfaces are probably weaker than 
the ionizing fluxes require to produce the WIM (\S \ref{sec:diving}). 
Also, the ionization state of the LIC cloud suggests that such clouds can 
account for only a small portion of the ionization associated with the 
WIM.  Nonetheless, there are several conclusions we can draw from existing 
observations and theory that are relevant to the WIM:

\begin{enumerate}

\item Hot-cool gas interfaces appear to be widespread through the 
interstellar medium. Wherever
interfaces with intermediate temperature gas ($T \sim 10^5 - 10^{5.5}$~K)
exist there will be ionizing radiation generated, producing diffuse 
ionization. The fact that the spectra peak in the EUV
and that a large fraction of the generated flux should be captured by the
cool cloud gas makes interface radiation a potentially significant source
of ionizing radiation in regions shielded from direct O-star Lyman
continuum photons. For example, if the interface is very convoluted, a
typical line of sight through a cloud may pass through many hot-cool
surfaces, resulting in a total \ha\ intensity for the cloud that is
comparable to what is observed for the HVCs.  Interface radiation
appears to be require to explain the properties of the local cloud near 
the sun.

\item The theoretical spectra from evaporating clouds and turbulent mixing 
layers are harder than that of the O~stars, which could result in the 
presence of some ions in the WIM that O~stars cannot produce.  

\item The type of interface that exists at the boundary of a cloud depends 
(at least) on the dynamics of the boundary, the geometry of the cloud and 
the degree to which thermal conduction occurs.  Thus refinements in theory 
and observations of good candidate clouds are needed to determine the 
extent to which interfaces contribute to ionization in our Galaxy and 
others.

\end{enumerate}

\section{Some questions for future study}
\label{sec:questions}

\subsection{What is the source of the elevated temeratures?}

Some of the line ratios, including \oiii/\ha\ and the largest observed [N 
{\sc ii}]/H$\alpha$ and [S {\sc ii}]/H$\alpha$ ratios do not seem to be 
explained by photoionization, even when hardening of the O~star's 
radiation is considered (\S \ref{sec:clumpy}).  Do shocks play a role 
\citep{cr01, Hidalgo05}?  Or is there some additional source of 
non-ionizing heat, such as photoelectric heating from interstellar dust 
particles or large molecules \citep{RC92, WD01}, dissipation of turbulence 
\citep{MS97}, magnetic reconnection \citep{raymond92}, shocks and cooling 
hot gas \citep{SSB93, cr01}?  Most of these mechanisms seem capable of 
generating heating rates of the order $10^{-26}\,{\rm erg}\,{\rm 
s}^{-1}\,{\rm cm}^{-3}$, the power requirement for the additional heat 
source \citep[see discussion in][]{RHT99}.  In addition, \citet{RHT99} 
show that for these non-ionizing sources, their heat can dominate over 
that from photoionization at low ($ \alt 10^{-1}$ cm$^{-3}$) densities, 
because their heating rates are proportional to the first power (or less) 
of the density, rather than the second power as with photoionization. This 
could explain the observed inverse correlation between the line ratios and 
the \ha\ intensity (see Fig.~\ref{fig:greglr}). Ways of discriminating 
between these mechanisms are very much needed, but none is yet 
forthcoming.

Can photoelectric heating be eliminated as a candidate for the 
supplemental heat?  In neutral gas, the photoejection of electrons from 
polycyclic aromatic hydrocarbon (PAH) molecules by stellar ultraviolet 
photons below the Lyman limit is believed to be a major source of 
heating.  However, PAHs may not be abundant in ionized gas, as suggested 
by both observational and theoretical considerations.  Observations and 
modeling of the Orion Nebula \hii\ region \citep[e.g.,][]{fer01}, show 
that PAHs are not present in the ionized gas there.  From a theoretical 
perspective, if PAHs were present in \hii\ regions their large Lyman 
continuum opacity \citep[e.g.,][]{WD01} would compete with 
hydrogen for ionizing photons and result in around 90\% of the ionizing 
photons being absorbed by PAHs \citep{mw2005}.  PAH opacity peaks 
around 17~eV, which, for example, would result in 
$\mathrm{He}^{+}$/$\mathrm{H}^{+}$ ratios in ionized gas containing PAHs 
being much higher than is observed in \hii\ regions surrounding late 
O stars and in the DIG.  On the other hand, silicaceous dust particles 
have a grayer wavelength dependent opacity and only absorb about 5\% of 
ionizing photons in the \citet{mw2005} simulations, not 
significantly changing the He$^+$ abundance or other line ratios.  
However, the role they may play in photoelectric heating is unclear.

\subsection{What is the spatial distribution of the gas?}

Does the wide-spread H$\alpha$ emission originate from a 
low density intercloud gas that occupies large volumes of the interstellar 
medium?  Or is some significant fraction of the H$\alpha$ coming from the 
denser ionized faces of clouds or superbubble walls?  Models investigating 
the different scenarios and comparisons of the resulting emission line and 
line ratio maps with observations, including electron column densities 
from pulsar dispersion measures may provide some answers.  
\citet{Hill08}, for example, have shown that the statistical distribution 
of emission measures in the WIM is matched very well by the density 
distribution produced in magnetohydrodynamic (MHD) models of an isothermal, mildly supersonic 
turbulent medium.  Also, new observational methods that sample the 
$\mathrm{H}^{+}$ at very small scales \citep{HSB03} or that access 
emission lines in a different spectral region \citep[e.g.,][]{KBM04}, may 
offer new insights about the small scale structure and dynamics of the gas 
as well as its larger scale distribution within the disk and halo.

\subsection{How much ionizing radiation escapes the galaxy?}

There have been several theoretical calculations of how much ionizing 
radiation escapes from galaxies \citep[e.g.,][]{ds94, wl, rs00, c02, 
CO02}.  The models show that if the interstellar medium is smooth the 
fraction of escaping photons is small.  Clumpy density structures allow 
for larger escape fractions through low density paths in the interstellar 
medium (see Fig. \ref{fig:wood1} above).  However, there are few 
observations to test these models and determine how much ionizing 
radiation actually escapes.  If interface radiation or shocks 
\citep{BSA07} are not a major source of ionization, then one promising 
avenue is to study the ionized surfaces of distant high velocity clouds 
surrounding the Galaxy \citep{Tea02, TRH98}.  \citet{bm99, p2003, TRH98} 
have all estimated that a few percent of the ionizing luminosity from the 
Galaxy would be required to explain the H$\alpha$ emission from high 
velocity clouds located $\sim$ 10~000 pc above the disk \citep{Wakker04}. 
As discussed in \S \ref{sec:clumpy}, the spectrum of the escaping 
radiation (e.g., from future observations of He~I/H$\alpha$ from 
intermediate and high velocity clouds) could reveal whether photons escape 
the galaxy through very low density channels or by filtering through 
density bounded ionized regions.

\subsection{Do hot, pre-white dwarf stars play a role?}

In their late stages of evolution, low-mass stars pass through a hot 
photospheric phase after shedding their outer envelopes. Once the stellar 
envelope (i.e., the planetary nebula) has expanded sufficiently and is 
optically thin to the ionizing radiation from the stellar core, the Lyman 
continuum photons are available to ionized the ambient interstellar 
medium. Such stars may be responsible for the ionization of small, 
localized regions in the low density interstellar medium \citep{RCM05}, 
but it is not known whether they also contribute significantly to the more 
diffuse ionizing radiation producing the WIM.  Their luminosities and 
their lifetimes in this phase are orders of magnitude smaller than that of 
massive O~stars; however, in comparison to O~stars, their numbers are 
enormous, and they are much more uniformly distributed throughout the the 
Galactic disk.  Early calculations by \citet{Hills72} indicated that the 
ionizing radiation from these hot pre-white dwarf stars could 
significantly influence the interstellar medium.  Since then, there has 
been progress in understanding the late stages of evolution of low mass 
stars, but little additional work has been carried out on their 
collective, large-scale influence on the interstellar medium.

\subsection{Is missing atomic data important?}

Presently photoionization models are unable to make accurate predictions 
of the [S {\sc ii}] emission from the DIG because the dielectronic 
recombination rates for sulfur are unknown \citep[see discusion 
in][]{ali91}. The determination of these rates is important because they 
impact the inferred S$^+$ and S$^{++}$ abundances in the DIG, and, because 
sulfur is an important coolant, influencing the predicted strengths of 
other lines. While efforts are underway to calculate these dielectronic 
recombination rates, there may also be a handle on the rates 
observationally via modeling the emission line strengths for large, low 
surface brightness \hii\ regions.

\subsection{What insights will new global models provide?}

A next step in modeling would be to test global dynamical models of the 
interstellar medium \citep[e.g.,][]{dab01, KLB07} to see whether their 
density structures can allow for O~star radiation to produce the observed 
ionization (H$\alpha$ emission) and temperature structure (line ratios) of 
the DIG.  This will be a formidable task, combining large-scale 3D 
dynamical and radiation transfer simulations.  A unique solution for {\it 
the} structure of the interstellar medium may not even be possible without 
also incorporating observations of the other gas phases.  However, 
combining dynamical and photoionization models would provide observational 
signatures that could be searched for.  Perhaps progress could be made in 
testing various scenarios---ruling out classes of models, determining 
what conditions (density and dynamics) are required to allow radiation to 
escape to the halo, and determining which models best fit the observed 
distribution and kinematics of the \ha\ over the sky.

While much progress has been made in understanding the nature of the warm 
ionized medium and the basic physical processes occurring within it, the 
future promises to be a very exciting time, when the advances in computing 
ability are combined with high spatial and spectral resolution 
observations of this gas and other major phases of the interstellar 
medium.  There is much still to be learned.

\begin{acknowledgments}
\label{ackn}

We thank Carl Heiles for his contributions, enthusiasm, encouragement, and 
insights over the years.  We also thank the meeting organizers, Y.-H. Chu 
and T. Troland, without whom this paper would not have been written. We 
thank two anonomous referees, whose comments and criticisms resulted in an 
improved paper. We also thank A. Ferguson, C. Hoopes, and J. Rossa for access to high-quality versions of their figures as well as their permission to include them in this review. We are also very grateful to our students and 
collaborators for their support and contributions.  RJR thanks Don Cox for 
many valuable conversations over the years regarding the nature of the 
WIM.  RJR \& LMH were supported by the National Science Foundation through 
grants AST 02-04973 and AST 06-07512, with assistance from the University 
of Wisconsin's Graduate School, Department of Astronomy, and Department of 
Physics. GJM acknowledges support from the University of Sydney 
Postdoctoral Fellowship Program and the National Science Foundation 
through AST 04-01416.  R-JD's work at Ruhr-University Bochum in this field 
is supported through DFG SFB 591 and through Deutsches Zentrum f\"ur Luft- 
und Raumfahrt through grant 50 OR 9707. JEB, AZ, \& CG acknowledge support 
through grants AYA2001--0435 from the Spanish Ministry of Science and 
Technology and AYA2004-08251-C02-01 from the Ministry of Education and 
Science. A. Zurita thanks the Consejer\'\i a de Educaci\'on y Ciencia de 
la Junta de Andaluc\'ia, Spain, for support.

\end{acknowledgments}

\section*{A list of acronyms and terms}
\label{aandt}

\begin{description}

\item{Chimney:} a superbubble that extends from the galactic midplane,
where it has been energized by supernovae, into the halo. Chimneys can 
be a conduit for hot gas and hydrogen ionizing radiation.

\item{DIG:} diffuse ionized gas characterized
by temperatures $\sim 10^4$~K and densities $\sim 10^{-1}$~cm$^{-3}$ that 
occupies the disk and halo of many spiral galaxies.  In our Galaxy the 
DIG is often referred to as the warm ionized medium or WIM.

\item{DM:} dispersion measure; the free electron density integrated along 
the line of sight to a pulsar; $\int^{P}_{0}$~$n_e$~ds. Because the 
free electron density n$_{e} \approx$ n$_{\rm{H}^+}$ in the interstellar 
medium, DM is essentially the column density of the $\mathrm{H}^{+}$.

\item{EM:} emission measure; the product of the proton density 
times the electron density integrated along the line of sight. In regions 
of nearly fully ionized hydrogen, emission measure is essentially 
$\int$~n$_{\rm{e}}^2$~ds.

\item{EPIC:} European Photon Imaging Cameras on the XMM-Newton's X-ray 
telescopes.

\item{FIR:} far infrared (wavelengths $\sim 10^2~\mu$m). 

\item{Forbidden line:} an emission line produced by the decay of an 
excited metastable state in an atom or ion, denoted by square brackets 
around the ion's symbol. In the low density WIM/DIG, the collisional 
de-excitation time scales are much longer than the radiative decay time 
scales, and therefore the intensity of a forbidden line measures the 
excitation rate of the metastable state due to collisions by the thermal 
electrons in the plasma.

\item{the Galaxy:} with a capital `G', our galaxy, the Milky Way.

\item{Galactic latitude:} The angular distance above the Galactic equator, 
the midplane of the Milky Way.

\item{\ha:} the emission at 6563~\AA\ produced by the hydrogen 
Balmer-alpha (n $= 3 \rightarrow 2$) transition following the 
recombination of ionized hydrogen.

\item{\hi\ region:} a cloud composed primarily of neutral hydrogen atoms.

\item{\hii\ region:} a discrete region of photoionized hydrogen associated 
with a hot star; a classical ``emission nebula'' or ``Str\"omgren 
Sphere'', as opposed to the more wide-spread, lower density 
$\mathrm{H}^{+}$ (the WIM, DIG) not clearly associated with a single, 
discrete source of ionization.

\item{$\mathrm{H}^{+}$:} ionized hydrogen; in this paper ionized hydrogen 
associated with the wide-spread, low density WIM/DIG.

\item{HST:} the Hubble Space Telescope

\item{HVC:} high velocity cloud; a neutral hydrogen cloud, usually located 
far from the Galactic disk and not partaking in the rotation of the disk.

\item{Ionization parameter:} the ratio of the ionizing photon density to 
the electron density, which along with the spectrum, determines the 
population of ionization states in photoionized gas.

\item{LIC:} the very local interstellar cloud within a few parsecs of the 
sun.

\item{LyC:} Lyman continuum; energies above the Lyman limit, the 
ionization potential of hydrogen (13.6 eV).

\item{Magellanic Stream:} an extended complex of neutral atomic hydrogen 
gas associated with a pair of satellite galaxies, the Large and 
Small Magellanic Clouds, orbiting the Galaxy.

\item{O and B~stars:} hot stars that emit Lyman continuum (LyC) photons.  
Massive O~stars are located in active star forming regions, emit a large 
fraction of their luminosity in the LyC, and are believed to be the 
primary source of ionization for the WIM/DIG.

\item{PAH:} polycyclic aromatic hydrocarbon; a large interstellar 
molecule that can inject heat into the interstellar via its ionization by 
ultraviolet photons from stars.

\item{R:} rayleigh; a unit of surface brightness for emission lines, equal 
to $10^6/4\pi$ photons s$^{-1}$ cm$^{-2}$ sr$^{-1}$.  An ionized region 
with an emission measure EM = 2.25 cm$^{-6}$~pc and a temperature of 
8000~K has an \ha\ surface brightness of 1~R.

\item{SFR:} star formation rate.

\item{Spiral galaxy:} a galaxy that has a flat, rotating disk of stars, 
gas, and dust.

\item{Superbubble:} a large cavity of hot ($\sim 10^6$~K), ionized gas 
created by the combined kinetic energy of multiple supernovae occurring 
within an active star formation region.

\item{$T_e$; $T_i$:} electron temperature and ion temperature, 
respectively. In general, $T_e$ = $T_i$ in the interstellar medium.

\item{TML:} turbulent mixing layer; the transition region between two 
adjacent parts of the interstellar medium that have very different 
temperatures and are moving with respect to each other.

\item{VLA:} The Very Large Array radio synthesis telescope near Socorro, 
New Mexico.

\item{WHAM:} the Wisconsin H-Alpha Mapper; a remotely controlled 
observatory dedicated to the detection and study of faint emission lines 
from the interstellar medium of the Galaxy.

\item{WIM:} the wide-spread, warm, ionized medium in the Galaxy 
characterized by temperatures $\sim 10^4$~K and densities $\sim 
10^{-1}$~cm$^{-3}$; sometimes also called the ``Reynolds Layer''.

\item{WNM:} wide-spread neutral atomic hydrogen characterized
by temperatures $\sim 10^3$~K and densities $\sim 10^{-1}$~cm$^{-3}$.

\item{XMM-Newton:} the Multi-Mirror Mission orbiting X-ray observatory 
of the European Space Agency.

\item{$z$:} the perpendicular distance from the midplane of a spiral galaxy.

\end{description}

\bibliography{heilesreview-final}

\end{document}